\documentclass[useAMS,usenatbib]{mn2e}

\usepackage{natbib}
\usepackage{graphicx}
\usepackage{graphics}
\usepackage[T1]{fontenc}
\usepackage{times}

\newcommand{\hms}[3]{\ensuremath{{#1}^\mathrm{h}{#2}^\mathrm{m}{#3}^\mathrm{s}}}
\newcommand{\dms}[3]{\ensuremath{{#1}^\mathrm{\circ}{#2}\mathrm{'}{#3}\mathrm{''}}}

\title[Origin of filaments in Centaurus\,A]{Filaments in the southern giant lobe of Centaurus\,A: constraints on nature and origin from modelling and GMRT observations}
\author[Sarka Wykes et al.]{Sarka Wykes$^{1,2}$\thanks{E-mail: sarka@astro.ru.nl}, Huib T. Intema$^{3,4}$, Martin J. Hardcastle$^{5}$, Abraham Achterberg$^{1}$, \newauthor Thomas W. Jones$^{6}$, Helmut Jerjen$^{7}$, Emanuela Orr\'u$^{1,8}$, Alex Lazarian$^{9}$, \newauthor Timothy W. Shimwell$^{10}$, Michael W. Wise$^{2,8}$ and Philipp P. Kronberg$^{11,12}$
\\
$^{1}$Department of Astrophysics/IMAPP, Radboud University Nijmegen, P.O. Box 9010, 6500 GL Nijmegen, The Netherlands\\
$^{2}$Anton Pannekoek Institute for Astronomy, University of Amsterdam, P.O. Box 94249, 1090 GE Amsterdam, The Netherlands\\
$^{3}$National Radio Astronomy Observatory, Socorro NM 87801, USA\\
$^{4}$National Radio Astronomy Observatory, 520 Edgemont Road, Charlottesville, VA 22903-2475, USA\\
$^{5}$School of Physics, Astronomy and Mathematics, University of Hertfordshire, College Lane, Hatfield, Hertfordshire AL10 9AB, UK\\
$^{6}$School of Physics and Astronomy and the Minnesota Supercomputing Institute, University of Minnesota, Minneapolis, NM 55455, USA\\
$^{7}$Research School of Astronomy \& Astrophysics, Australian National University, Mount Stromlo Observatory, Cotter Road, Weston\\ Creek, ACT 2611, Australia\\
$^{8}$ASTRON, P.O. Box 2, 7990 AA Dwingeloo, The Netherlands\\ 
$^{9}$Department of Astronomy, University of Wisconsin, 475 North Charter Street, Madison, WI 53706, USA\\
$^{10}$CSIRO Australia Telescope National Facility, PO Box 76, Epping NSW 1710, Australia\\ 
$^{11}$Los Alamos National Laboratory M.S. B283, Los Alamos NM 87545, USA\\
$^{12}$Department of Physics, University of Toronto, 60 St. George Street, Toronto, M5S 1A7, Canada
}
\begin{document}

\date{Accepted 2014 May 22. Received 2014 April 24; in original form 2014 February 11}

\pagerange{\pageref{firstpage}--\pageref{lastpage}} \pubyear{2014}

\maketitle

\label{firstpage}

\begin{abstract}
We present results from imaging of the radio filaments in the southern giant lobe of Centaurus\,A using data from GMRT observations at 325 and 235\,MHz, and outcomes from filament modelling. The observations reveal a rich filamentary structure, largely matching the morphology at 1.4\,GHz. We find no clear connection of the filaments to the jet. We seek to constrain the nature and origin of the {\it vertex} and {\it vortex} filaments associated with the lobe and their role in high-energy particle acceleration. We deduce that these filaments are at most mildly overpressured with respect to the global lobe plasma showing no evidence of large-scale efficient Fermi\,I-type particle acceleration, and persist for $\sim2-3$\,Myr. We demonstrate that the dwarf galaxy KK\,196 (AM\,1318--444) cannot account for the features, and that surface plasma instabilities, the internal sausage mode and radiative instabilities are highly unlikely. An internal tearing instability and the kink mode are allowed within the observational and growth time constraints and could develop in parallel on different physical scales. We interpret the origin of the {\it vertex} and {\it vortex} filaments in terms of weak shocks from transonic MHD turbulence or from a moderately recent jet activity of the parent AGN, or an interplay of both.
\end{abstract}

\begin{keywords}
galaxies: individual (Centaurus\,A) -- galaxies: jets -- instabilities -- radio continuum: galaxies -- techniques: image processing -- turbulence.
\end{keywords}

\section{Introduction} \label{sect:introduction}
From studies of both high- and low-power radio galaxies over the past three decades, considerable observational evidence has emerged for inhomogeneous, filamentary lobes, e.g., Cygnus\,A \citep{PER84}, 3C\,310 \citep{BRE84}, Hercules\,A \citep{DRE84, GIZ03}, Fornax\,A \citep{FOM89}, Pictor\,A \citep{PER97}, 3C\,353 \citep{SWA98}, M\,87 \citep{OWE00, FOR07}, NGC\,193 \citep{LAI11}, B2\,0755+37 \citep{LAI11}, with the bulk of the observations being conducted with the Very Large Array (VLA) in the GHz regime. The filamentarity has implications for the internal structure of the lobes. However, no consensus exists on whether the magnetic field in the lobes has a low filling factor and electrons are uniformly distributed, or the electron population tracks the magnetic field enhancements closely. Positionally varying magnetic field strength was claimed for lobes of a number of Fanaroff-Riley class\,II (FR\,II) \citep{FAN74} sources (e.g. \citealp{HAR05, GOO08}), in contrast with the western giant lobe of the source Fornax\,A\footnote{Morphologically, Fornax\,A is FR\,II class by the original \citep{FAN74} definition; in terms of luminosity, it is on the boundary FR\,I/FR\,II.}, for which a positionally varying electron energy spectrum is favoured \citep{SET11}. 
\begin{figure*}
\includegraphics[width=1.0\textwidth]{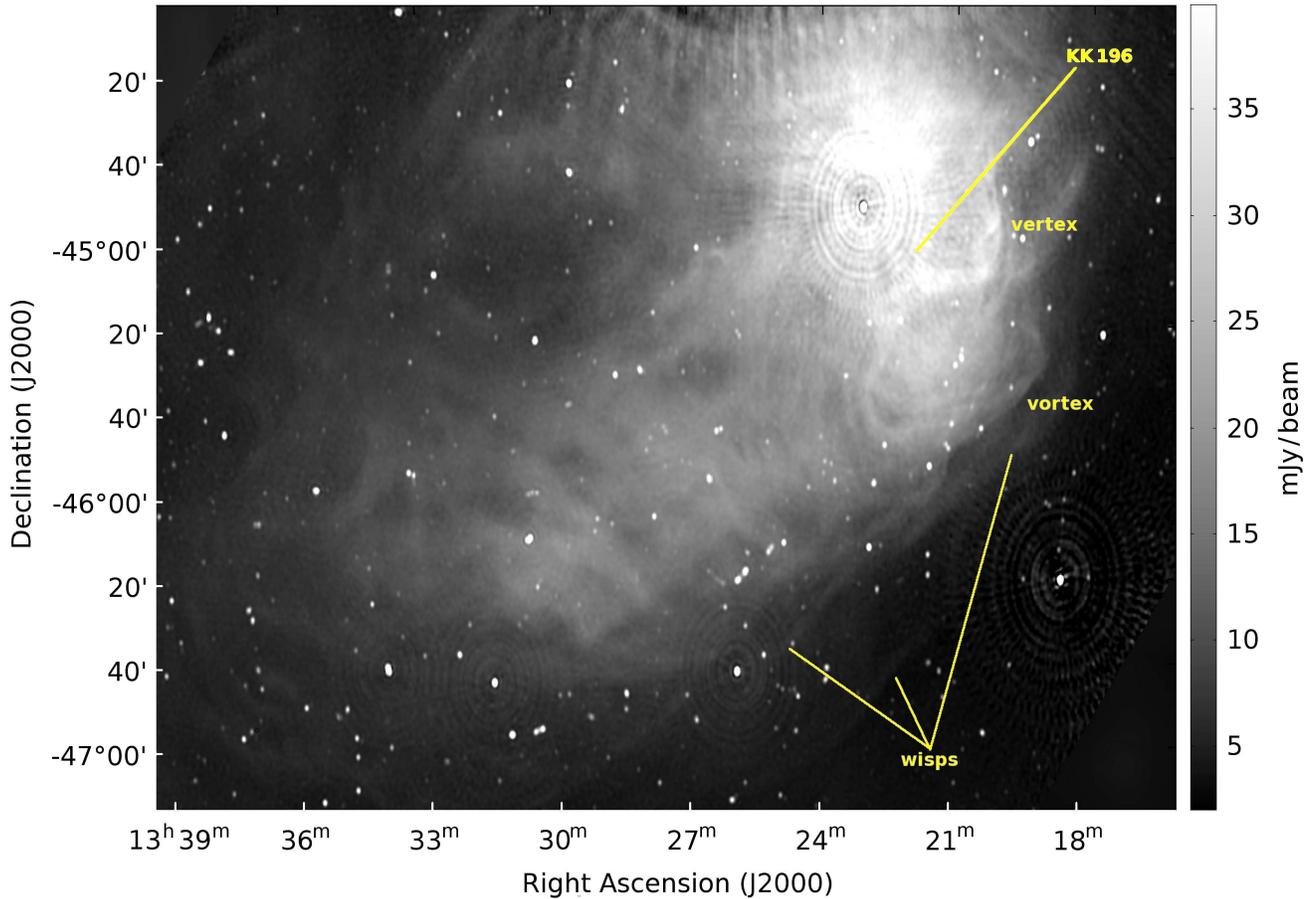}
\caption{Combined ATCA and Parkes 1.4\,GHz continuum image at $60\times40$\,arcsec angular resolution of the southern giant lobe of Centaurus\,A, indicating the {\it vertex} and {\it vortex} filaments and the position of the dwarf galaxy KK\,196. The radio galaxy core and the inner lobes are beyond the north edge of the image. The artefact protruding to the northern part of the giant lobe originates from the bright inner lobes. The artefact centred on RA\,$\hms{13}{23}{04.2}$, DEC\,$\dms{-44}{52}{33.3}$ is the background quasar PKS\,1320--446 at a redshift of $z\sim1.95$, the artefact on RA\,$\hms{13}{18}{30.02}$, DEC\,$\dms{-46}{20}{35.2}$ the background quasar MRC\,1315--460 (PMN\,J1318--4620) of $z\sim1.12$ and the artefact on RA\,$\hms{13}{19}{21.59}$, DEC\,$\dms{-44}{36}{46.7}$ the background source MRC\,1316--443. Adapted from \citet{FEA11}.} 
\label{fig:fig1}
\end{figure*}

Filamentary structure does not necessarily imply turbulence, but magnetohydrodynamical (MHD) turbulence implies filamentary structure in synchrotron emission (e.g. \citealp{EIL89, HAR13, WYK13}). MHD turbulence amplifies and transports magnetic fields which in turn control lobe viscosity, conductivity and resistivity, as well as the acceleration and propagation of cosmic rays (e.g. \citealp{LEE03, JON11}). The presence of turbulence might in some cases be akin to a development of plasma instabilities. Various types of instabilities, promoting growth of filament-like features, could develop inside or on the surface of a radio lobe. Hydrodynamical (HD) instabilities, such as Kelvin-Helmholtz (KH), Rayleigh-Taylor (RT) and Richtmyer-Meshkov (RM) are relevant since they can lead to flow patterns that naturally filament and can amplify ambient magnetic fields (e.g. \citealp{JUN95, RYU00}). MHD instabilities such as the resistive tearing instability, the sausage mode and the kink mode are also apposite, as are radiative instabilities. 

Utilising the Australia Telescope Compact Array (ATCA) and the $64$\,m Parkes telescope for imaging at 1.4\,GHz with $49$\,arcsec angular resolution, \cite{FEA11} have discovered intricate filamentary features associated with the northern and southern giant lobes of Centaurus\,A (Fig.\,\ref{fig:fig1}). Centaurus\,A is the nearest (3.8$\,\pm\,$0.1\,Mpc; \citealp{HARR10})\footnote{At that distance, $1$\,arcmin corresponds to $1.1$\,kpc.} Fanaroff-Riley class\,I (FR\,I) radio galaxy, hosted by the massive elliptical galaxy NGC\,5128. Due to its luminosity and proximity, Centaurus\,A is an outstanding testbed for models of jet energetics, particle acceleration, and the evolution of low-power radio galaxies in general. Centaurus\,A's northern jet (angular size $\sim4.0$\,arcmin) and its immediate surroundings, the bright inner lobes ($\sim5.5$\,arcmin each) and the northern middle lobe ($\sim33$\,arcmin) have been extensively studied (e.g. \citealp{TIN98, MOR99, HAR03, KRA03, HAR06, CRO09, KRA09, MUL11, NEF14}; Israel et al., in preparation). However, Centaurus\,A's proximity to Earth has hampered for a long time comprehensive investigations of its giant (i.e. outer) lobes ($\sim4.3^{\circ}$ each), whose substructure has been mapped only recently in the aforementioned work by \cite{FEA11}.

Topics of great current interest are the ages of the giant lobes and the lobe particle content and pressure. \cite{HAR09} and \cite{YAN12} have determined radiative ages of Centaurus\,A's giant lobes: the former obtaining $\sim30$\,Myr based on synchrotron ageing fitting the single-injection Jaffe-Perola model \citep{JAF73}, the latter $\la80$\,Myr reasoning that ages significantly larger than a few tens of Myr are not consistent with the observations of gamma-ray inverse-Compton emission. The above values would imply that the giant lobe front ends expand at respectively $\sim0.030$ and $\ga0.011c$, i.e. faster than Centaurus\,A's inner lobes ($\sim0.009c$, \citealp{CRO09}), in discord with expectations. The dynamical age calculations by \cite{WYK13} give $\sim560$\,Myr based on buoyancy arguments, and the estimates by \cite{EIL14} give $\sim500$\,Myr$-1.5$\,Gyr relying on dynamical models of the growth of the giant lobes. Giant lobe thermal electron content evaluations also show an inconsistent picture: $n_{\rm e,th}\sim1\times10^{-4}$\,cm$^{-3}$ gauged independently from X-ray and radio observations \citep{STA13, SUL13} versus $n_{\rm e,th}\sim5.4\times10^{-9}$\,cm$^{-3}$ based on entrainment calculations \citep{WYK13}. The preceding value, which is similar to the thermal content of the Centaurus\,A intragroup medium, $n_{\rm th}\sim1\times10^{-4}$\,cm$^{-3}$ \citep{SUL13, EIL14}, would make Centaurus\,A exceptional among lobed radio sources, which normally show cavities associated with the lobe (e.g. \citealp{BIR04, WIS07, CAV10}) implying that the internal densities of the lobes are below those of the intragroup/intracluster medium. Giant lobe pressure estimates vary from $p_{\rm th}\sim8.0\times10^{-14}$\,dyn\,cm$^{-2}$ \citep{STA13, SUL13, STE13} and $p_{\rm th}\sim9.0\times10^{-14}$\,dyn\,cm$^{-2}$ \citep{FRA14} which are close to the minimum pressure of the lobes, to $p_{\rm th}\sim3.2\times10^{-13}$\,dyn\,cm$^{-2}$ \citep{EIL14} and $p_{\rm th}\sim1.5\times10^{-12}$\,dyn\,cm$^{-2}$ \citep{WYK13}. 

\cite{HAR09}, \cite{SUL09}, \cite{SUL11}, \cite{WYK13} and \cite{EIL14} have considered MHD turbulence in Centaurus\,A's giant lobes. \cite{WYK13} have argued for mildly sub-Alfv\'enic turbulence in those lobes which allows for the existence of relatively long-lived filaments\footnote{MHD simulations (e.g. \citealp{JON11}) indicate that this is also true for trans-Alfv\'enic and mildly super-Alfv\'enic turbulence. Filaments last roughly an eddy turnover time for the scale of the eddies that stretch them.}. A detailed description of the ensemble of the filaments has been offered by \cite{FEA11}, and \cite{STA13} and \cite{WYK13} have drawn some attention to the prominent filamentary features in the southern giant lobe, the {\it vertex} and the {\it vortex} (see Figs.\,\ref{fig:fig1}\,--\,\ref{fig:fig4}), whose nature and origin are as yet ill-constrained. The {\it vertex} (Largest Angular Size, LAS, $\sim\!34$\,kpc), at $\sim\!2^{\circ}$ from the core and $\sim0.5^{\circ}$ from the position of the background point source PKS\,1320--446, is slightly curved and shows variations in surface brightness. The {\it vortex} (LAS $\sim\!53$\,kpc), at about $\sim2.5^{\circ}$ from the core and lying immediately interior to the western part of the lobe, has been likened \citep{FEA11} to the mushroom-shaped structure seen in the eastern giant lobe of the FR\,I radio source M\,87 \citep{OWE00}. \cite{FEA11} have proposed a number of explanations to be origin of the filaments in Centaurus\,A: the {\it vertex} and {\it vortex} might derive from an enhanced jet activity, from KH instabilities at the lobe-intragroup interface, or from the passage of the dwarf irregular galaxy KK\,196 (AM\,1318--444), a Centaurus group member at $3.98\pm0.29$\,Mpc \citep{JER00, KAR07}, through the lobe. 

To gain more insight into physical processes occuring in radio galaxies's lobes, it is essential to have access to multifrequency observations, including very-low frequency bands. We have chosen to use the Giant Metrewave Radio Telescope (GMRT) to study the properties of the {\it vertex} and {\it vortex} filaments at 325, 235 and 150\,MHz. At these frequencies, the field of view of the GMRT is large enough to fit the combined {\it vertex}\,--\,{\it vortex} region in one or two pointings. The combination of short and long baselines (ranging from $\sim\,100$\,m up to $\sim\,25$\,km) enables the detection of, and separation between, large structures of the angular size of the filaments and compact radio sources along the same line of sight. The GMRT has no baselines that sample the largest scales of radio emission from the giant lobes, which are therefore naturally surpressed. Besides the intrinsic delicacies of handling low-frequency radio data and imaging diffuse, extended radio emission, processing these observations is particularly challenging. As observed from GMRT, the {\it vertex} and {\it vortex} filaments are at low elevation (always $<26^{\circ}$), which impacts on the $uv$-coverage and ionospheric air mass in an unfavourable way. Also, the large brightness of the background quasar PKS\,1320--446 and of Centaurus\,A's inner lobes (e.g. Fig.\,\ref{fig:fig1}) can cause serious dynamic range (DR) limitations in the entire image.

The remainder of the paper is organized as follows. Section\,\ref{sect:obs} describes the GMRT observations and data reduction. In Section\,\ref{sect:results}, we present the new GMRT images and combine the GMRT data with those of the ATCA at higher frequencies with the aim of establishing spectral indices of the {\it vertex} and {\it vortex} filaments. We discuss filament pressure, ageing and particle acceleration constraints, and expound upon turbulence properties on different scales of the giant lobes in Section\,\ref{sect:interpretation}. We then test various scenarios for the origin of the {\it vertex} and {\it vortex} and show that these structures are most likely not unique among the lobe filaments and are plausibly identified with weak shocks in transonic MHD turbulence. The key findings are summarised in Section\,\ref{sect:summary}.

Throughout the paper, we use J\,2000.0 coordinates, and define the energy spectral indices $\alpha$ in the sense $S_{\!\nu}\propto\nu^{-\alpha}$.

\section{Observations and data reduction} \label{sect:obs}

GMRT observations of the {\it vertex} and {\it vortex} filaments at 325, 235 and 150\,MHz were carried out in 2012 May and 2013 February (project codes 22\_038 and 23\_060). Due to the low elevation of the target, observations were limited to $<5$\,hours per night, requiring a total of $8$\,nights. A journal of these observations is given in Table\,\ref{tab:journal}. Visibilities for two polarizations (RR and LL) were recorded in spectral line mode to enable narrow-band RFI excision and prevent bandwidth smearing. For calibration purposes we observed two standard calibrators: 3C\,286 as the (primary) flux density and bandpass calibrator, observed for $20$\,min at the start and/or end of each run, and 3C\,283 as the (secondary) phase calibrator, observed for $5$\,min every $50$\,min. During observations, the standard GMRT pointing error correction model was applied on both targets, as well as 3C\,286.

\begin{table*}
\begin{center}
\caption{Journal of GMRT observations.}
\label{tab:journal}
\begin{tabular}{llllllll}
\hline \\ [-1.5ex]
Observing dates & Central obs. freq. & Primary beam & Time on target & Working antennas & Integration time & Bandwidth & Freq. channels\\
  & (MHz) & (arcmin) &  ({\it vertex}, {\it vortex}) (h) & (total number) & (s) & (MHz) &\\
\multicolumn{8}{l}{} \\
\hline \hline \\ [-1.5ex]
$27-28$ May $2012$ & $323$ & $81$  & $3.7$, $4.0$ & $29$, $27$ & $4$ & $33.3$ & $256$\\
$24-27$ Feb $2013$ & $234$ & $114$ & $8.4$, $8.1$ & $28$, $28$ & $8$ & $16.7$ & $256$\\
$22-23$ Feb $2013$ & $148$ & $186$ & $8.2$        & $29$       & $8$ & $16.7$ & $256$\\
[1ex] \hline
\end{tabular} \\
\end{center}
\end{table*}
The data\footnote{Long Time Accumulation (LTA) -- the native data format from the GMRT correlator -- were written out to FITS format and imported into Astronomical Image Processing System ({\sc aips}) and Common Astronomy Software Applications ({\sc casa}).} reduction was conducted using {\sc aips} (version {\tt 31DEC12}; \citealp{GRE03}), the Python-based extension {\sc spam} (\citealp{INT09, INT14}), and the {\tt clean} imaging task in {\sc casa} (version {\tt 4.1.0}; \citealp{MCM07}) to create the final low-resolution maps. {\sc spam} was used to correct for (direction-dependent) ionospheric phase corruptions, which can be a dominant source of error at sub-GHz frequencies. Overall, ionospheric conditions were relatively quiet during all observing nights, as judged from the slowly varying gain phases. For all observations, except 2012 May 28, flux and bandpass calibration were derived from 3C\,286, adopting the modified Perley--Taylor flux model as described by \citet{INT11}. For this, we excluded the shortest (central square) baselines, and performed some manual flagging of obviously bad antennas/times/polarizations based on gain calibration tables. Calibration results were then applied to the target field data. In the 2012 May 28 observation, a combination of factors rendered the data on 3C\,286 unusable. In this case, we have used 3C\,283 as primary calibrator, adopting a 325\,MHz flux density of $23.3$\,Jy (as derived from the 2012 May 27 observations). To account for uncertainties in the flux scale transfer from 3C\,286 to 3C\,283, and for the absence of the pointing error correction for 3C\,283, we adopt an additional $10$ percent uncertainty in the flux scale of the {\it vortex} image at 325\,MHz. To reduce the target field data volume, the RR and LL polarizations \citep{CON69} were combined into Stokes $I$, every $8-10$ channels were averaged to form $25-30$ wider frequency channels, and time resolution was averaged down to $16$\,seconds.

\subsection{Calibration and imaging of compact emission} \label{sect:calim}

The target field data was initially calibrated and imaged at `high resolution only' by excluding visibilities from the inner $1$\,k$\lambda$ of the $uv$-plane and choosing a weighting scheme between robust and uniform ({\tt robust} $=-1$ in the {\sc aips} convention). 
\begin{table*}
\begin{center}
\caption{Properties of GMRT maps per region and frequency.}
\label{tab:properties}
\begin{tabular}{lllll}
\hline \\ [-1.5ex]
Image & Central obs. freq. & Angular resolution & Position angle & Central rms noise \\
      & (MHz)              & (arcsec)           &  (degree)      & (mJy\,beam$^{-1}$) \\
\multicolumn{5}{l}{} \\
\hline \hline \\ [-1.5ex]
{\it vertex} high-resolution & $323$ & $16.3\times6.9$ & $10.7$ & $0.30$\\ 
{\it vortex} high-resolution & $323$ & $15.9\times6.9$ & $11.6$ & $0.23$\\ 
{\it vertex} high-resolution & $234$ & $22.5\times9.0$ & $0.6$ & $0.80$\\ 
{\it vortex} high-resolution & $234$ & $23.0\times9.6$ & $-0.5$ & $0.76$\\ 
{\it vertex/vortex} high-resolution & $148$ & $42.1\times12.6$ & $-9.3$ & $3.78$\\ 
{\it vertex/vortex} low-resolution & $323$ & $66.5\times38.0$ & $17.4$ & $3.8$\\ 
{\it vertex/vortex} low-resolution & $234$ & $60.2\times41.3$ & $-5.1$ & $10.8$\\ 
[1ex] \hline
\scriptsize{Note. Rms values corrected for $T_{\rm sys}$.}
\end{tabular} \\
\end{center}
\end{table*}
With this approach, we can obtain good gain calibrations for all antennas without having to deal with the large-scale emission immediately. Indeed, the same calibration can be used later to image the large-scale emission (see Section\,\ref{sect:large}). Each target field was initially phase-calibrated against a simple $\sim10$~point source model derived from the SUMSS 843\,MHz catalogue \citep{MAU03}, followed by wide-field (facet-based) imaging and CLEAN deconvolution. In all imaging, we automatically put tight CLEAN boxes encompassing emission peaks (above five times the central noise level) to guide the deconvolution and suppress CLEAN bias. Within the CLEAN boxes, emission was CLEANed down to twice the central noise level. Extra facets were added at the locations of known bright outlier sources within $4$ primary beam radii. This was followed by two rounds of phase-only self-calibration, and one round of amplitude and phase self-calibration. In between rounds, bad data was removed by flagging spurious points in the gain calibration tables and residual visibility amplitudes.

Self-calibration was followed by two rounds of ({\sc spam}) ionospheric calibration and imaging. This includes peeling \citep{NOO04} of $10-20$ of the brightest sources. The latter was most useful to reduce the DR-limiting effects of the point source PKS\,1320--446. Outside the main beam, we found that at 325\,MHz the GMRT primary beam strongly attenuated the inner lobes of Centaurus\,A, and also the background FR\,I radio galaxy PKS\,B1318--434, so that they do not put any DR limitations on our image quality. At 235\,MHz, the inner lobes caused slight ripples across our image, but these were reduced by peeling. At 150\,MHz, the inner lobes strongly affected the entire field of view, rendering the image inutile for our purpose of detecting faint diffuse emission. Many attempts (including peeling) to improve this failed. We will disregard the 150\,MHz observations in what follows. Table\,\ref{tab:properties} lists the properties of the final high-resolution images.

\subsection{Imaging of large-scale emission} \label{sect:large}

With gain calibration sorted out during the high-resolution imaging, we have explored several ways of imaging the highly-resolved diffuse components that are embedded in the southern giant lobe of Centaurus\,A. To maximize our signal-to-noise on the {\it vertex} and {\it vortex}, we converged on a method in which we used the {\sc casa} imager task {\tt clean} in mosaicking mode, simultaneously imaging and CLEAN-ing all visibilities of the two pointings into one final image per frequency. We were restrained in the use of multi-scale deconvolution, because it is as yet unsupported in mosaicking mode.

Before importing the visibilities into {\sc casa}, we have pre-subtracted (in {\sc aips}\,/\,{\sc spam}) the CLEAN components (of $\geq5$ times the background rms noise) of all point sources found in the compact emission image from the visibility data sets, while temporarily applying the appropriate direction-dependent gain calibrations. Furthermore, since {\sc spam} functionality is not yet available in {\sc casa}, we have used a single gain table to calibrate each data set, namely the one corresponding to the field centre. This lack of direction-dependent ionospheric calibration seems to have limited effect while imaging using solely the shortest baselines.

For imaging the diffuse emission, we used a Gaussian weight taper, suppressing visibilities from baselines longer than $2.5$\,k$\lambda$. The few visibilities from baselines shorter than $150\,\lambda$ were excluded during imaging to (i) remove strong ripples coming from the very few high-amplitude visibilities that sense (but completely undersample) the largest-scale structures, and (ii) roughly match our data to the ATCA interferometric data that is part of the study by \citet{FEA11}, and which is also used in our analysis.

\begin{figure*}
\includegraphics[width=\linewidth]{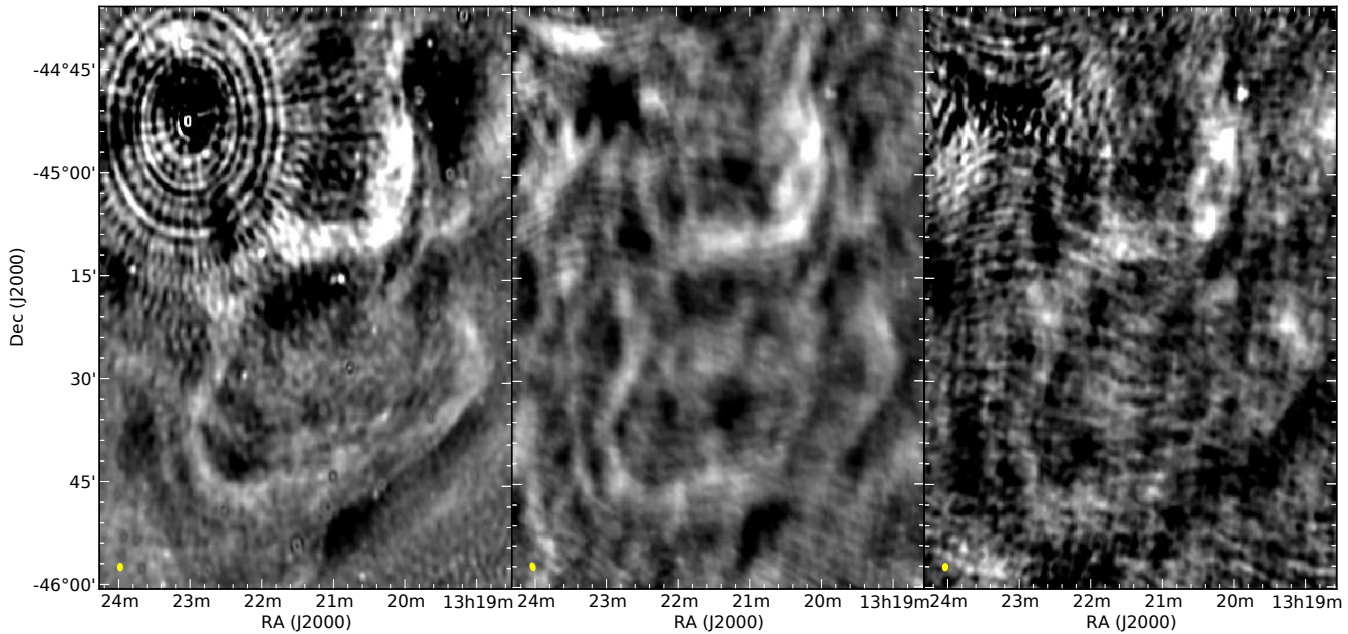}
\caption{From left to right: ATCA 1.4\,GHz, GMRT 325\,MHz and GMRT 235\,MHz continuum maps of the large-scale emission in the {\it vertex} and {\it vortex} fields. The beam size is indicated by the ellipse in the bottom left-hand corner. The rms noise of the GMRT images is given in Table\,\ref{tab:properties}.} 
\label{fig:fig2}
\end{figure*}

The final GMRT low-resolution images (Fig.\,\ref{fig:fig2}, middle and right panel) have a DR (defined here as the ratio of peak flux to the off-source rms away from the image centre) of $10^2:1$. We determine the rms noise level ($1\sigma$) in those images as $\sim3.8$\,mJy\,beam$^{-1}$ (325\,MHz) and $\sim10.8$\,mJy\,beam$^{-1}$ (235\,MHz) (see Table\,\ref{tab:properties}).

\subsection{Flux density levels} \label{sect:levels}

Large-scale radio emission that is resolved out by the GMRT interferometer, but is still registered by individual GMRT antennas, enters the system as sky noise, routinely expressed as sky temperature $T_{\rm sky}$. At low radio frequencies, the sky noise can make a significant contribution to the total system temperature $T_{\rm sys}$, which also includes receiver noise, ground radiation and potentially other, smaller caches of noise. GMRT observations are by default not corrected for variations in $T_{\rm sys}$ when moving the telescope across the sky (e.g. \citealp{TAS07,SIR09, INT11}). This causes a target field flux scale error $T_{\rm sys,target} / T_{\rm sys,fluxcal}$ when transferring the gain calibration from flux calibrator to target field.

In our case, the large-scale emission of the southern giant lobe of Centaurus\,A is filling most (or all) of our target fields of view at all observing frequencies, resulting in considerably higher sky temperatures than in the field of our calibrators 3C\,286 and 3C\,283, leading to large flux scale errors. For correcting the resulting flux scale error of our target fields we rely on the model discussed essentially by \cite{SIR09}, with frequency dependence also incorporated. The resultant $T_{\rm sys}$ values, which are based on the duration of our observations, are codified in Table\,\ref{tab:tsys}. Instantaneous values of $T_{\rm sky}$ and $T_{\rm receiver}$ independently are not meaningful for our purpose and were not computed. All gain ratios and correction factors were calculated with regard to our flux density calibrator 3C\,286. The errors on the flux correction factors amount to $\sim4$ percent. Note that systematics in calibration while imaging will dominate the flux density calibration at these frequencies with the GMRT which are in the range $\sim10-12$ percent. Thus, the total error on the flux density is $\sim15$ percent.

\begin{table}
\begin{center}
\caption{Flux scale corrections, based on $T_{\rm sys}$ estimates in the context of the model described by Sirothia (2009).}
\label{tab:tsys}
\begin{tabular}{lll}
\hline \\ [-1.5ex]
Pointing & $T_{\rm sys}$ & Flux correction factor\\
& (K) & \\
\multicolumn{3}{l}{} \\
\hline \hline \\ [-1.5ex]
3C\,286 at 323\,MHz        & $123.57$ & $1.0$\\
3C\,286 at 234\,MHz        & $222.96$ & $1.0$\\
3C\,283 at 323\,MHz        & $129.28$ & $1.05$\\
3C\,283 at 234\,MHz        & $235.17$ & $1.05$\\
{\it Vertex} at 323\,MHz   & $248.78$ & $2.01$\\
{\it Vertex} at 234\,MHz   & $469.93$ & $2.11$\\
{\it Vortex} at 323\,MHz   & $227.07$ & $1.84$\\
{\it Vortex} at 234\,MHz   & $437.15$ & $1.96$\\
{\it Vertex}/{\it vortex} at 325\,MHz & $239.68$ & $1.93$\\
{\it Vertex}/{\it vortex} at 235\,MHz & $455.56$ & $2.04$\\
[1ex] \hline
\end{tabular} \\
\begin{tabular}{l}
\scriptsize{Note. The $T_{\rm sys}$ values are based on a model for our duration of the observations.}
\end{tabular}
\end{center}
\end{table}

Since the final 325\,MHz and 235\,MHz low-resolution images have been produced with visibility data from both pointings combined, we assign single flux correction factors per frequency. Based on Table\,\ref{tab:tsys}, we adopt respectively $1.93$ at 325\,MHz and $2.04$ at 235\,MHz.

\subsection{1.4\,GHz image processing} \label{sect:14processing}

An overview of the ATCA and Parkes observations and imaging of the giant lobes at 1.4\,GHz can be found in \cite{FEA09, FEA11}. We have used a {\it vertex}/{\it vortex} ATCA data subset from the above work and subtracted the background sources from the image plane using the source extraction tool {\sc p}{\scriptsize Y}{\sc bdsm}\footnote{\tt http://dl.dropboxusercontent.com/u/1948170/html/\\index.html}. This map is used later in the article for spectral index extraction (Sections\,\ref{sect:flux} and \ref{sect:index}) and turbulence modelling (Section\,\ref{sect:turbulence}).
\begin{table*}
\begin{center}
\caption{Measured flux density, largest angular size and diameter (over the minor axis) of the {\it vertex} and {\it vortex} filaments at 1.4\,GHz and at 325 and 235\,MHz.}
\label{tab:parameters}
\begin{tabular}{llllll}
\hline \\ [-1.5ex]
Filament & $S_{\rm 1.4\,GHz}$ & $S_{\rm 325\,MHz}$ & $S_{\rm 235\,MHz}$ & LAS & Diameter \\
         & (Jy)               & (Jy)               & (Jy)               & (arcmin) & (arcmin) \\
\multicolumn{6}{l}{} \\
\hline \hline \\ [-1.5ex]
{\it Vertex} & $1.25\pm0.15$ & $4.90\pm0.24$ & $4.61\pm0.73$ & $31$ & $7$ \\
{\it Vortex} & $0.72\pm0.14$ & $2.60\pm0.40$ & $3.05\pm0.60$  & $58$ & $4$ \\
[1ex] \hline
\scriptsize{Note. Flux density corrected for background.}
\end{tabular} \\
\end{center}
\end{table*}

\section{Results} \label{sect:results}

\subsection{Radio morphology and filament flux densities} \label{sect:flux}

As is apparent from Fig.\,\ref{fig:fig2}, the {\it vertex}, {\it vortex} and other bright filaments maintain their coherence over the frequency range 1.4\,GHz\,--\,235\,MHz, i.e., the 1.4\,GHz emission appears spatially closely associated with the emission at 325\,--\,235\,MHz. Given the better sampling of large-scale structures at low frequencies, auxiliary filamentary structure could have been revealed by our observations, if present. We find some additional filamentary features but no clear indication in the GMRT images for the {\it vertex}/{\it vortex} and other features being connected to the core/jet. 
\begin{figure}
\includegraphics[width=\linewidth]{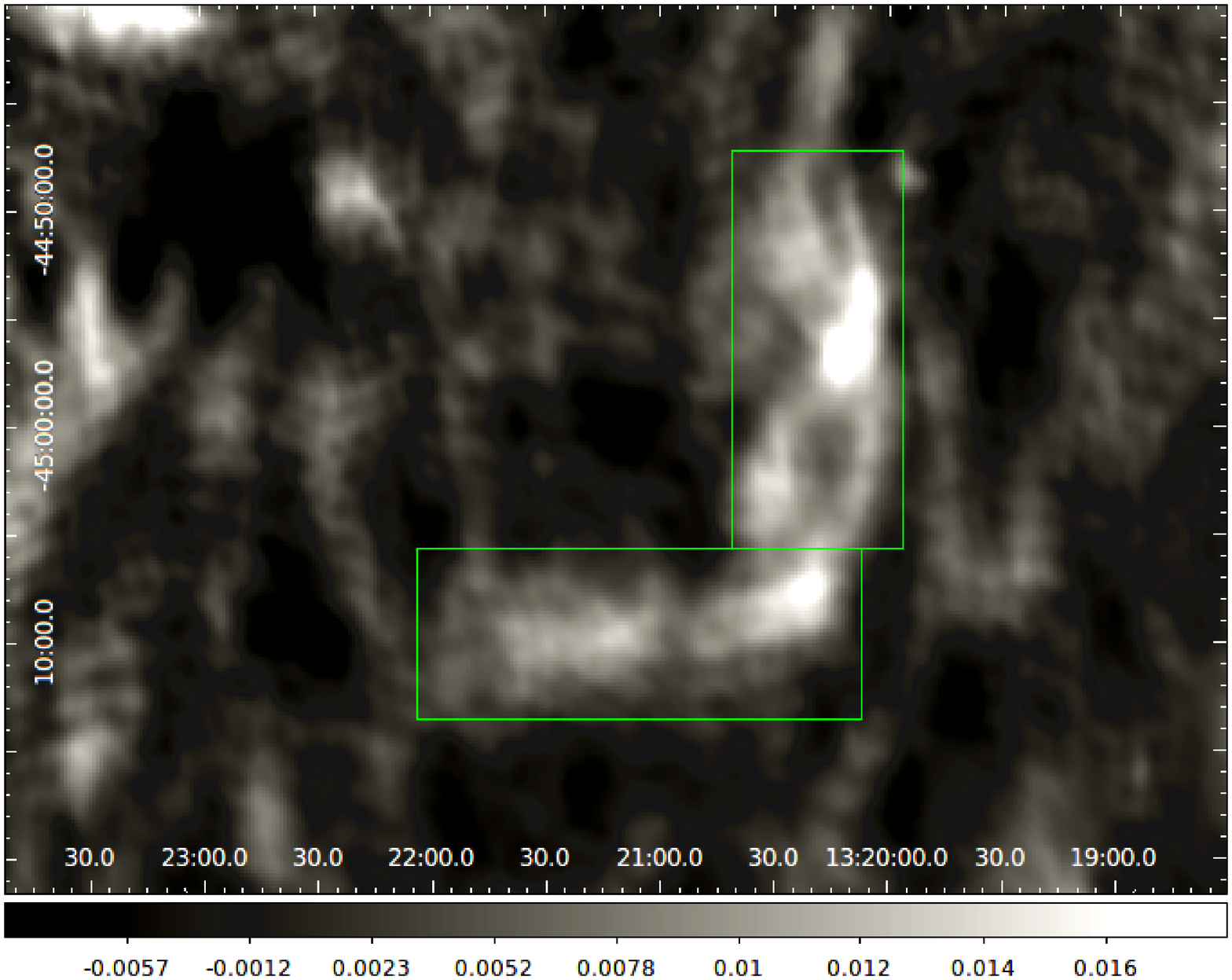}
\caption{GMRT 325\,MHz map of the {\it vertex} filament with spectral extraction regions used for Table\,\ref{tab:parameters} overlaid. The scale is in Jy\,beam$^{-1}$.} \label{fig:fig3}
\end{figure}
\begin{figure}
\includegraphics[width=\linewidth]{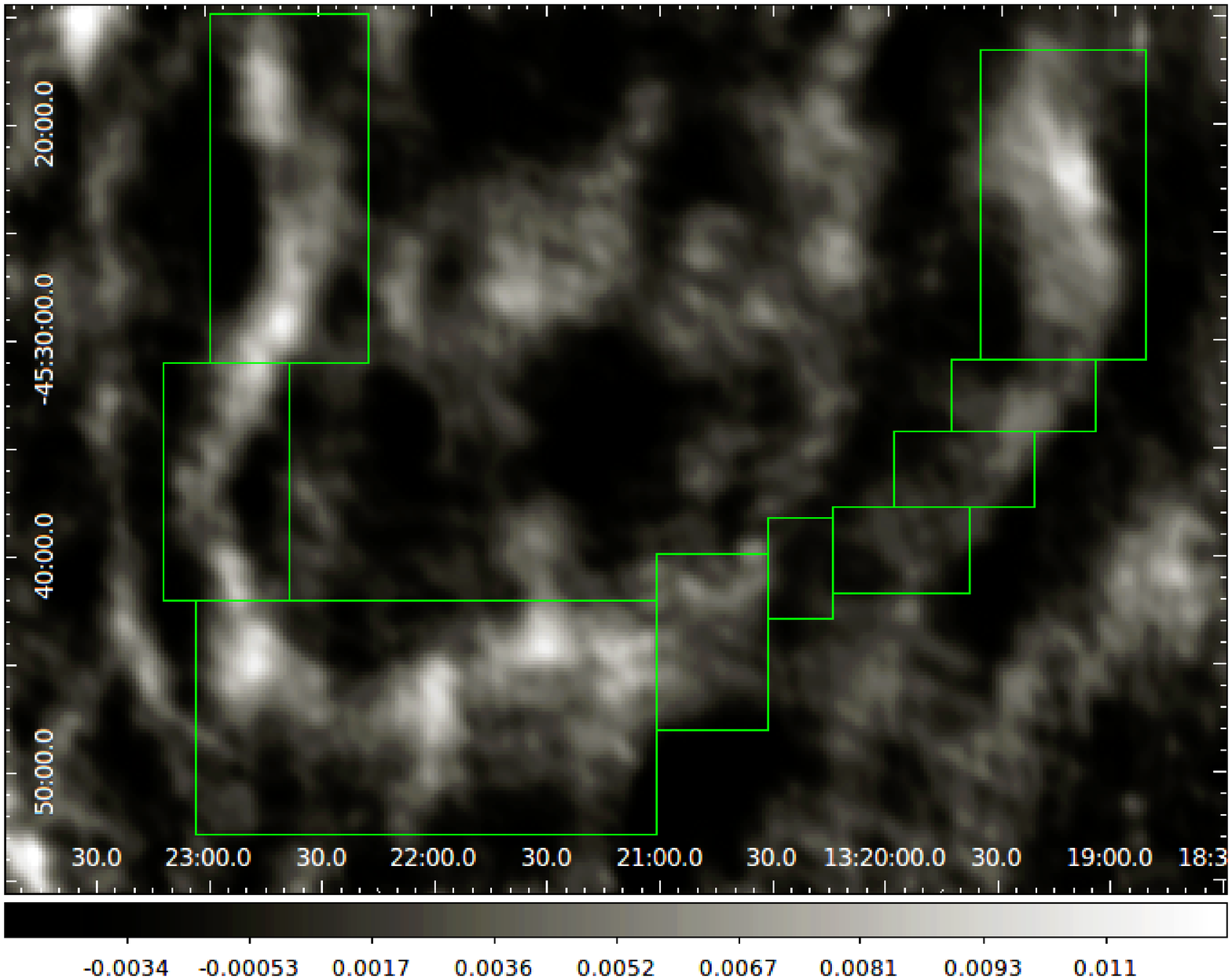}
\caption{GMRT 325\,MHz map of the {\it vortex} filament with spectral extraction regions used for Table\,\ref{tab:parameters} overlaid. The scale is in Jy\,beam$^{-1}$.} \label{fig:fig4}
\end{figure}The topology of the filamentary mesh is unexplored; however, the {\it vertex} filament seems twisted, bright at what may be intersections, and its northern part appears to be built up of two segments which themselves potentially contain substructure in form of approximately parallel threads (most clearly visible in Fig.\,\ref{fig:fig3}). At all available frequencies, the {\it vortex} morphology resembles a fleshy fungus; the eastern `cap of the mushroom's head' is displaced by about $4$\,arcmin to the south as we go to lower frequencies. Pronounced in the 325\,MHz image (Fig.\,\ref{fig:fig2}, middle panel) is a short bow-like filament north-east of the {\it vertex}, which is `behind' the interferometric rings caused by the PKS\,1320--446 source in the 1.4\,GHz image (Fig.\,\ref{fig:fig2}, left panel). Fainter, but still well discernible (in Fig.\,\ref{fig:fig2} left and middle, and Fig.\,\ref{fig:fig3}) is a pipe-like filament close to and west of the {\it vertex}.

We have measured the {\it vertex} and {\it vortex} flux densities from FITS files making use of {\sc ds{\footnotesize 9}}\footnote{\tt https://hea-www.harvard.edu/RD/ds9/site/Home.html} and the Funtools\footnote{\tt https://www.cfa.harvard.edu/{\scriptsize$\sim$}john/funtools} library and defining rectangular regions encompassing the filaments (see Figs.\,\ref{fig:fig3} and \ref{fig:fig4}): 2 regions of respectively 5217 and 5828 pixels for the {\it vertex}, and 9 regions (to account for its curvature, and for a slightly different position of its south-east part as a function of frequency) for respectively 2310, 1519, 504, 912, 819, 800, 3956, 4268 and 8320 pixels for the {\it vortex} (on a synthesised beam area of 28.65 pixels at 325\,MHz and 28.20 pixels at 235\,MHz). Equally-sized rectangular regions were used elsewhere on the maps, well away from obvious outliers (positive and negative) and from edge-effects, for background determination for which we effected $15$ trials. We handled
the standard deviation of the background values to estimate the error on the source regions.

Table\,\ref{tab:parameters} lists the final, background-corrected {\it vertex} and {\it vortex} flux densities. The 1.4\,GHz value is an order of magnitude less than the flux density of these filaments quoted in \cite{FEA11}; however, there they represent the combined ATCA+Parkes flux densities at 1.4\,GHz and are not background subtracted.

\subsection{Filament spectral index} \label{sect:index}

We have utilised our GMRT 325\,--\,235\,MHz radio continuum images and the existing ATCA 1.4\,GHz continuum image to determine the spectral index. To this end, we have matched the shortest baselines, and have disregarded visibilities from baselines shorter than $0.15$\,k$\lambda$. We have fitted power laws in frequency to the flux densities and errors in Table\,\ref{tab:parameters} augmented with the errors on the calibration in the GMRT data, minimizing $\chi^2$. From our analysis, the resultant best-fitting spectral indices are $\alpha=0.81\pm0.10$ ({\it vertex}) and $\alpha=0.83\pm0.16$ ({\it vortex}).

Given the large errors on the spectral index for the entirety of the individual filaments, we cannot determine variations of the spectral index along them (i.e., a spectral index analysis for filament subregions would be meaningless).

\section{Interpretation} \label{sect:interpretation}

\subsection{Vertex and vortex: the nature} \label{sect:nature}

A detailed radio spectral energy distribution (SED) analysis can provide lower limits on the ages of the filaments and clues about the filament's nature and origin. The radio SED of distinct filaments in a handful of lobes where such measurements are accessible shows a slightly flatter spectrum compared to that of the general lobe plasma, e.g. $\alpha_{\rm 1.3\,GHz}^{\rm 8.4\,GHz}\sim1.1$ vs. $\alpha_{\rm 1.3\,GHz}^{\rm 8.4\,GHz}\sim1.6$ in the FR\,I/II source Hercules\,A \citep{GIZ99}, which might indicate either more recent particle acceleration or magnetic field enhancement at the filament \citep{TRI94}. More specifically, the spectral indices of the ring-like filaments in Hercules\,A obtained by spectral tomography \citep{GIZ03} are in the range $\alpha_{\rm 4.8\,GHz}^{\rm 8.4\,GHz}\sim 0.70-1.15$, and at lower GHz-frequencies in the range $\alpha_{\rm 1.3\,GHz}^{\rm 4.8\,GHz}\sim0.70-0.90$, with a general trend of flatter spectral indices at lower frequencies. The {\it arc} filament in the western lobe of Hercules\,A, comparable in morphology and orientation to the {\it vertex} and {\it vortex}, shows a significantly steeper index than any other filamentary feature in Hercules\,A's lobes: $\alpha_{\rm 4.8\,GHz}^{\rm 8.4\,GHz}=0.90\pm0.05$ and $\alpha_{\rm 1.3\,GHz}^{\rm 4.8\,GHz}=1.14\pm0.04$ (see their table 5): this is surprising if it is associated with a recently shocked region. Given that Hercules\,A is probably in a driving phase, more useful comparisons with Centaurus\,A might include FR\,I sources such as Hydra\,A and 3C\,310, or the source Fornax\,A. However, no spectral index measurements of individual filaments within the lobe volume of these sources (and any other FR\,I or FR\,I/II radio galaxy) are yet at hand.

Centaurus\,A's integrated southern giant lobe diffuse emission shows a spectral index $\alpha_{\rm 1.4\,GHz}^{\rm 5.0\,GHz}=0.55\pm0.02$ and $\alpha_{\rm 408\,MHz}^{\rm 1.4\,GHz}=0.47\pm0.06$ \citep{HAR09} and $\alpha_{\rm 118\,MHz}^{\rm 1.4\,GHz}=0.63 \pm 0.01$ \citep{MCK13} in the region of the {\it vertex}. Around the location of the {\it vortex}, the measured spectral indices are $\alpha_{\rm 1.4\,GHz}^{\rm 5.0\,GHz}=0.71\pm0.05$ and $\alpha_{\rm 408\,MHz}^{\rm 1.4\,GHz}=0.62 \pm 0.12$ \citep{HAR09} and $\alpha_{\rm 118\,MHz}^{\rm 1.4\,GHz}=0.65 \pm 0.01$ \citep{MCK13} although this region does not encompass the {\it vortex} in its entirety. Various authors (\citealp{FEA11, STA13, STE13, WYK13, EIL14}) have modelled (parts of) the giant lobes as recently undergoing particle re-acceleration; this idea is congruous with the rather flat spectral indices of the global lobe plasma. An important point is that the spectral indices that we measure for the filaments (Section\,\ref{sect:index}) are not significantly flatter than these lobe spectral indices; instead, they are steeper ($90$ percent confidence level) than or similar to (given the relatively large errors on the GMRT flux densities) the global lobe plasma values measured by \cite{MCK13}. This places constraints on the possible character of the filaments that we discuss in more detail in Sections\,\ref{sect:acceleration} and \ref{sect:ages}.

A number of theoretical works have suggested that a correlation between $B$-field and particle density may be weak, essentially because $B$-field fluctuations relate to Alfv\'enic turbulence which does not compress the plasma, and also on grounds of the so-called `reconnection diffusion' which violates the flux-freezing condition (see for more in-depth discussions, e.g., \citealp{PAS03, CHO03, SAN10}). This is supported by X-ray observations of radio lobes (e.g. Pictor\,A, \citealp{HAR05}; 3C\,353, \citealp{GOO08}) where the inverse-Compton emission does not follow the pattern seen in radio-synchrotron emission from the lobes. We discuss the possible behaviour of the magnetic field in the filaments in Sections\,\ref{sect:pressure} and \ref{sect:ages}.

It has thus far been unclear whether the {\it vertex} and {\it vortex} filaments are overpressured as argued for M\,87 \citep{HIN89, FOR07} and for Pictor\,A \citep{PER97}, or in pressure equilibrium with the medium within which they are embedded. The large pressure jump factors associated with the {\it vertex} and {\it vortex} ($p_2/p_1\sim30$ for the {\it vertex} and $\sim240$ for the {\it vortex} for the heat capacity ratio of $\gamma=5/3$, and $p_2/p_1\sim15$ for the {\it vertex} and $\sim80$ for the {\it vortex} for $\gamma=4/3$) presented by \cite{FEA11} on grounds of an idealised case of an initially spherical cocoon collapsing into a torus after a passage of a strong shock, would point towards large filament overpressure, and therefore to fairly flat filament spectral indices if these are set by particle acceleration. For $p_2/p_1\sim30$ and $p_2/p_1\sim240$ pressure ratios and $\gamma=5/3$, the Mach number ($\mathcal{M}$) from the Rankine-Hugoniot jump relations is respectively $\mathcal{M}\sim4.9$ and $\mathcal{M}\sim13.9$; for $p_2/p_1\sim15$ and $p_2/p_1\sim80$ pressure ratios and $\gamma=4/3$, it is respectively $\mathcal{M}\sim3.6$ and $\mathcal{M}\sim8.4$. However, it is not clear whether such large overpressure factors are consistent with the radio observations. We will elaborate on pressure considerations in Sections\,\ref{sect:pressure} and \ref{sect:jet}.

\subsubsection{Filament pressure} \label{sect:pressure}

In this section we consider the pressure in the filaments and its implications for their dynamics.

Since the radio emission from the {\it vertex} and {\it vortex} is synchrotron radiation, we can calculate the pressure of these filaments from their flux densities and estimated volumes if we make some assumptions about the relative energy densities in $B$-field, electrons, and non-radiating particles. Ideally, we would use the lowest radio frequency at our disposal for the flux density measurement, i.e., 150\,MHz, to minimize the effect of any age-related steepening in the electron spectrum. However, because the 150\,MHz data have defied imaging attempts (see Section\,\ref{sect:calim}) and the 235\,MHz data possess a relatively large error on the flux, we resort to the flux densities at 325\,MHz. We estimate the volumes of the {\it vertex} and {\it vortex}, treating them as cylinders with length $l=48$\,kpc and radius $r=3.2$\,kpc ({\it vertex}) and $l=105$\,kpc and $r=1.8$\,kpc ({\it vortex}), to be respectively $V\sim4.5\times10^{67}$ and $V\sim3.1\times10^{67}$\,cm$^3$.

One possible approach is to determine the minimum energy in the lobes and filaments. To do this we adopt the \cite{MYE85} minimum-electron-energy model and perform the minimum-energy calculations numerically using the code of \cite{HAR98}. This yields for the {\it vertex} an equipartition $B$-field strength of $B\sim2.4$\,$\mu$G and a minimum pressure of $p_{\rm min}=2U_B/3\sim1.5\times10^{-13}$\,dyn\,cm$^{-2}$ while for the {\it vortex} we find an equipartition field $2.2\,\mu$G and $p_{\rm min}\sim1.3\times10^{-13}$\,dyn\,cm$^{-2}$. Compared to the global minimum pressure of the lobes of $p_{\rm min}\sim4.5\times10^{-14}$\,dyn\,cm$^{-2}$ (based on results by \citealp{HAR09} building as well on the \citealp{MYE85} formalism), these are a factor $\sim3$ higher\footnote{For a sheet-like filament, the volume would be smaller; the emissivity would then have to be larger to produce the flux density (which is fixed), so $p_{\rm min}$ would increase, but only by a relatively small factor.}. Such an overpressure could be identified with weak shocks, $\mathcal{M}\sim1.7$ ({\it vertex}) and $\mathcal{M}\sim1.6$ ({\it vortex}), from the Rankine-Hugoniot jump relations.

However, in the model by \cite{WYK13}, the giant lobes are clearly not in minimum pressure; the dominant pressure component is provided by thermal material, with $p_{\rm th}\sim1.5\times10^{-12}$\,dyn\,cm$^{-2}$, with the pressures due to relativistic electrons and $B$-field being much lower. In this situation it is not clear what we should assume for the pressures in the various components in the filaments. We consider one possible limiting case, in which the thermal and relativistic electron pressures remain constant in the filaments while the $B$-field strength increases to provide the observed increase in synchrotron emissivity. We find that, by keeping the relativistic electron number density $n_{\rm e,rel}$ fixed, the $B$-field has to increase by a factor $\sim6.5$ ({\it vertex}) and factor $\sim6.0$ ({\it vortex}) with respect to the ambient $B$-field (for which we adopted $0.9$\,$\mu$G from \citealp{ABD10}), to produce the radio-bright filaments. We obtain $U_{B\rm+e}\sim1.7\times10^{-12}$\,dyn\,cm$^{-2}$ and thus $p_{B\rm+e}\sim5.8\times10^{-13}$\,dyn\,cm$^{-2}$ ({\it vertex}), and $U_{B\rm+e}\sim1.2\times10^{-12}$\,dyn\,cm$^{-2}$ and thus $p_{B\rm+e}\sim3.9\times10^{-13}$\,dyn\,cm$^{-2}$ ({\it vortex}). In this case, the filaments would be only mildly overpressured (a factor $\sim1.3$) with respect to the medium in the giant lobes, which translates to $\mathcal{M}\sim1.0$. (We note that as this is a limiting case, the overpressure factor could be even lower than this if both electron and $B$-field energy densities are varied to produce the filament emissivity.) 

The $B$-field estimated above, although strong compared to the field estimated from inverse-Compton observations in the giant lobes as a whole \citep{ABD10}, is not energetically dominant: with a thermal pressure of $p_{\rm th}\sim1.5\times10^{-12}$\,dyn\,cm$^{-2}$ \citep{WYK13}, the filaments would be $B$-field dominated for a filament strength $B\ga6\,\mu$G. However, this conclusion depends strongly on our assumption about the thermal pressure in the giant lobes. With $p_{\rm th}\sim3.2\times10^{-13}$\,dyn\,cm$^{-2}$ \citep{EIL14}, they would already be $B$-field dominated for $B\ga3\,\mu$G. In the less likely case of giant lobe pressure close to the minimum pressure, the $B$-field limit would be lower. If the $B$-fields are amplified through turbulence, a hard upper bound to the strength of the filament $B$-field could be placed in terms of a balance between Maxwell and Reynolds stresses (i.e. balance between the magnetic tension and turbulence stresses, which is the saturation regime), leading to $B^2/(8\pi)=0.5\rho v_{\rm t}^2$, with $\rho$ the mass density and $v_{\rm t}$ the turbulent speed. We have no means of independently calculating $v_{\rm t}$ (see also the remarks in Section\,\ref{sect:turbulence}), hence we cannot obtain a maximum attainable $B$-field of the filament via this route.

The above results are inconsistent with the cocoon collapsing scenario suggested by \cite{FEA11}, which implied $\mathcal{M}\gg1.7$ (see the foregoing section); furthermore, the results make models relying on high-energy particle acceleration by the Fermi\,I mechanism in the giant lobes (as per, e.g., \citealp{PEE12} and \citealp{FRA14}) unlikely (see also Section\,\ref{sect:acceleration}).

\subsubsection{Particle confinement and ageing} \label{sect:acceleration}

As noted in Section\,\ref{sect:nature}, we find a somewhat steeper spectrum in the filaments than in the global lobe plasma. The steeper index is hard to explain in any model, but if it is real, it rules out models in which the excess emissivity in the filament region is related to (any type of) particle acceleration. Instead we need to consider models in which the electrons in the filaments have cooled more rapidly than those in the giant lobes. Probably the only viable excess-loss scenario involves particles confined in the filaments and so ageing faster \citep{TRI93}; hence in this section we consider whether confinement in the filaments can take place for relevant timescales.

To see whether electrons of the required energies (in the middle of the available observing frequencies of 235\,MHz to 2.2\,GHz (T. Shimwell, private communication, for the latter), i.e. $\sim1.2$\,GHz) stay confined in the {\it vertex} filament, we estimate the diffusion time:
\begin{equation}
\tau_{\rm diff} \simeq \xi\,r^2/r_{\rm g,e}\,c \,, \label{eq:diffus}
\end{equation}
where $\xi$ is a fudge factor for which we use $\xi\sim10$ (see \citealp{WYK13} and references therein). Assuming the $B$-field is parallel to the long axis of the filament (see Section\,\ref{sect:instabilities}), and taking the {\it vertex} radius $r\sim3.2$\,kpc, $B\sim5\,\mu$G and electron gyroradius $r_{\rm g,e}$ corresponding to the above energy of $\gamma\sim9\times10^3$, we get $\tau_{\rm diff}\sim330$\,Tyr. So for cross-field diffusion\footnote{There is potential for field line wandering and other mechanisms enhancing somewhat particle diffusion; our estimate might be considered a hard upper bound on the diffusion time.}, the electrons could remain in the filaments for a long time relative to the loss timescale. Note however that we assume that the filaments have field aligned along their length everywhere so as to imply cross-field diffusion for motion perpendicular to the filament. If we also assume effective pitch angle scattering, then the electrons can stream along the length of the filaments (and, presumably, off the end) on a timescale $l/c \sim0.1$\,Myr, which is much shorter than the loss timescale; moreover, unaged electrons must stream in at the same rate\footnote{In the presence of MHD turbulence, isolation of energetic particles and thus streaming of electrons at the speed of light over kpc-scale distances is unlikely, therefore the quoted value should be considered a hard lower limit.}. If there is some component of the field perpendicular to the filament long axis, then the timescale to move along the filament can become longer, but the timescale to move across it becomes shorter. So the confinement timescale is likely to be in the range $\sim0.1$\,Myr\,--\,$300$\,Tyr, and for a substantial fraction of that range {\it in situ} losses may be significant, thus allowing the electrons in the filaments to be more aged than those around them. However, we emphasize that the lifetime of the filaments themselves places a hard upper bound on the possible excess ageing of electrons trapped in them. We return to this point in the following section.

\subsubsection{Spectral ages} \label{sect:ages}

We now attempt to put constraints on the spectral (radiative) ages of the electrons in the filaments. Previously, \cite{HAR09} derived a synchrotron age of $24\pm1$\,Myr for Centaurus\,A's southern giant lobe region around the {\it vertex} and $29\pm1$\,Myr for the region around the {\it vortex}. It is clear that, naively, the steep spectra of the filaments would imply spectral ages older than those of the lobes, which is hard to understand, as noted in the previous section. In this section we explore whether excess-loss models of the sort described above can quantitatively explain the observed filament spectra. We use an injection index of $0.5$ to be consistent with \cite{HAR09}\footnote{It is plausible that the injection index is slightly steeper (as indicated for FR\,I sources by, e.g., \citealp{YOU05, LAI13}); the exact value of the injection index does not have a strong effect on our conclusions.}.

In this type of model we need to consider two $B$-fields: $B_{\rm present}$, i.e. the field that the electrons are currently in, and $B_{\rm age}$, i.e. the one that they have undergone most of their ageing in (see in this context the preceding section). $B_{\rm age}$ is an emission-weighted average of all the $B$-fields the electron has been in since acceleration; it is possible for it to be lower, equal to or higher than $B_{\rm present}$, depending on the stage of the filament evolution. For a fixed break frequency, the synchrotron age is
\begin{equation}
t_{\rm sync} \propto \frac{B_{\rm present}^{1/2}}{B_{\rm age}^2 + B_{\rm CMB}^2}\,,
\end{equation}
where $B_{\rm CMB}$ is the equivalent magnetic field strength giving the cosmic microwave background (CMB) energy density, and represents inverse-Compton losses. If $B_{\rm present}>B_{\rm age}$, e.g. if the electrons had spent most of their radiative lifetimes in a low-field region, this would make the filament spectral age exceed the spectral age of the lobe by an even larger factor. If we increase $B_{\rm age}$ (relative to the rest of the lobe) we can in principle obtain a filament spectral age below that of the lobe; however, $B_{\rm age}$ has to increase substantially because its equipartition value is significantly below $B_{\rm CMB}$ (which is about $3.3\,\mu$G). In practice, though, a high value of $B_{\rm age}$ could only be maintained on a timescale comparable to the dynamical lifetime of the filament, which we estimate (see Section\,\ref{sect:turbulence}) to be at most 3\,Myr. Forcing the spectral age to be as low as this requires extreme values of $B_{\rm age}$: fixing $B_{\rm present}$ to its maximum plausible value of $5.5\,\mu$G (see Section\,\ref{sect:pressure}) we obtain $B_{\rm  age}\sim30\,\mu$G, and even with $B_{\rm present}\sim0.9\,\mu$G as in the giant lobes at present (i.e. allowing an increase in the electron number density to produce the observed increased emissivity of the filaments) we require $B_{\rm age}\sim18\,\mu$G. These $B$-field strengths seem very high. Assuming that electrons are confined in the filament as discussed above, one can envisage a `squeezed state' followed by a `relaxed state' (i.e. recoil), then during the squeeze the electrons in the filament are ageing faster; so when the system relaxes back to the ambient $B$-field strength, the electrons in the filament are `older' than the electrons around them, and exhibit a steeper spectrum. However, in this picture very high magnetic fields must have been present at the peak of the squeezing.

We conclude that there is no very convincing explanation of the steep spectra of the filaments, if it is real, in the range of models we have considered. We emphasize that, because of the lack of a broad frequency range, our spectral age fit is not very robust, and more data are needed to confirm our picture. In addition, we have assumed a single electron population in the lobes. There could, in fact, be a second electron population (not to be confused with `secondary electrons' from proton-proton or proton-photon collisions), from two or more episodes of jet activity with some time in between for radiative/adiabatic losses; this would elevate the emission at the low-frequency end of the range and might be made visible in the filaments by elevated $B$-field strengths there. However, there is hardly any spectral evidence for this second population either in Centaurus\,A itself or in other FR\,I and FR\,II sources, and so we regard this explanation as speculative without confirmation from low-frequency observations of the giant lobes.

\subsubsection{Magnetic twist?} \label{sect:twist}

Both the ATCA 1.4\,GHz and the low-frequency GMRT images seem to indicate a twisting of the {\it vertex} filament. In a flux tube with a quasi-cylindrical geometry and with axial field $B_{\rm z}$ and azimuthal field $B_{\rm \phi}$, the amount of twist of a field line follows from the relation d$x/B_{\rm z} = r\,$d$ \phi/B_{\rm \phi}$ along a given field line. If the radius of the tube is $r$ and its length $l$, the total amount of twist of a field line at the edge of the tube is obtained from (e.g. \citealp{PRI94})
\begin{equation}
\Delta \phi = \frac{l\,B_{\rm \phi}(r)}{r\,B_{\rm z}}\,.
\label{eq:twist}
\end{equation}
We have observational constraints on the filament length and width (see Table\,\ref{tab:parameters} and Section\,\ref{sect:pressure}). If the twist is $\sim2\pi$, as seems from the available radio images, and hence the length over which the twist exists is $\sim31$\,kpc, then the ratio $B_{\rm \phi}/B_{\rm z}\sim0.5$. We will return to this in Section\,\ref{sect:instabilities}.

In the simple case, where the cylindrical filament carries a constant current density
over its cross section and for constant $B_{\rm z}$, the static MHD equilibrium gives the gas pressure on the tube's axis as $p\,(r=0) = p\,(r) + B_{\rm \phi}^2(r)/4\pi$. From this and Equation\,\ref{eq:twist} follows that the observed amount of twist $\Delta \phi$ constrains $B_{\rm \phi}$, and thereby the pressure difference: using $B_{\rm \phi}\sim(r\,\Delta \phi/l)B_{\rm z}$ we can write
\begin{equation}
\frac{\Delta p}{p}=\frac{(r\,\Delta \phi/l)^2}{1+(r\,\Delta \phi/l)^2}\,\frac{B^2}{4\pi p}\,.
\end{equation}
Thus, if the twisting scenario is correct, for {\it plasma}\,$\beta = p/(B^2/8\pi)\sim1$ (as is likely for the brightest filaments, see Section\,\ref{sect:pressure}) and $r\,\Delta \phi/l\sim0.5$, and assuming that $B_{\rm z}$ is constant inside and at the edge of the filament, the {\it vertex} cannot be compressed by a large factor and hence not greatly overpressured (i.e. compatible with strong shocks) with respect to the interfilament plasma.

\subsection{Vertex and vortex: the origin} \label{sect:origin}

In this section we put forward, as the most attractive origins of the {\it vertex} and {\it vortex} (i) temporary enhancements of the activity of the extant or extinct jet leading to squeezing of the plasma in the southern giant lobe, and (ii) internal MHD turbulence. \cite{FEA11} have likewise suggested jet activity, and additionally (iii) surface instabilities related to turbulence and (iv) the passage of the dwarf galaxy KK\,196 through the lobe. We perform a `sanity check' of the turbulent properties of the giant lobes, and we investigate whether the lobe plasma is prone to a development of HD and MHD instabilities.

\subsubsection{Jet activity} \label{sect:jet}

The idea that the {\it vertex} and {\it vortex} filaments might originate from an episode (or episodes) of (increased) jet activity and possibly represent weak shocks arises from, jointly, their 1.4\,GHz and 325\,--\,235\,MHz morphology\footnote{The morphology alone can easily be generated in simulations of isotropic turbulence.} and their orientation (see also \citealp{FEA11}). The {\it vertex} and {\it vortex} may connect to the jet/core (as appears in the radio continuum images of, e.g., M\,87; but note that weak, propagating shocks from the jet/core do not inevitably connect to it) whereas turbulence-induced filaments are not expected to be anchored (to one another or anything else). At 1.4\,GHz, 325\,MHz and 235\,MHz, no connection to the current jet has been observed. For a filament age of about $2-3$\,Myr (see Section\,\ref{sect:turbulence}) it is not obvious that the filaments could originate from direct activity of the current jet if we believe its physical age (discussed by \citealp{WYK13}) is $\sim2$\,Myr. If the filaments result from variations in the core/jet activity and represent subsequent outbursts, the brighter of these, the {\it vertex}, ought to be spectrally and dynamically younger than the {\it vortex}, since it is smaller in size and positioned closer to the core. Arguments involving filament diameter in cases where the filaments appear as (half) rings have been used to compare filament ages in Hercules\,A and 3C\,310 \citep{MOR96}. In the scenario in which the {\it vertex} and {\it vortex} filaments originate from a relatively recent (enhanced) direct jet activity, it would seem natural for them to exhibit a `leading edge' and a `trailing edge', but this is not seen at either 1.4\,GHz or the GMRT frequencies. In this scenario, and if the cooling time of the electrons were short enough, we would also expect to see a spectral gradient along a path from upstream to downstream, which is not seen either. In any case, it is not plausible that Centaurus\,A's jet (current jet power $\sim1\times10^{43}$\,erg\,s$^{-1}$, pre-existing jet power probably $\sim1-5\times10^{43}$\,erg\,s$^{-1}$; see \citealp{WYK13}) would produce these filaments as either terminal shocks or as a bow shock, in this source. We additionally note that angular momentum from the jet rotation on small scales may add left-over rotational component to the lobes on large scales and possibly help to explain a twisting of the {\it vertex}; however the spectral index gradient would then unlikely be as clear cut for this filament.

If the filaments are slightly higher pressure than the surroundings and are energetically plasma-dominated (rather than $B$-field dominated), they expand with a velocity $v_{\rm exp}\simeq(\Delta p/\rho)^{1/2}$, where $\Delta p=p_{\rm int}-p_{\rm ext}$ is the difference between the filament and external pressure and $\rho$ is the mass density (taken to be similar inside and outside the filament). For a filament width $d$ the expansion timescale is $t_{\rm exp}=d/v_{\rm exp}\simeq (p/\Delta p)^{1/2}(d/c_{\rm s})$, with $c_{\rm s}=\sqrt{\gamma k T/\mu m_{\rm H}}$ the adiabatic sound speed. In what follows we assume $\Delta p/p={\cal O}(1)$, so that the expansion time scale satisfies $t_{\rm exp}\simeq t_{\rm cs}$, the sound crossing time, and we estimate the timescale for expansion considering the timescale for sound to cross (the shortest dimension of) the structures:
\begin{equation}
t_{\rm cs}=d/c_{\rm s}\,.
\end{equation}
Using the heat capacity ratio $\gamma=5/3$, the mean particle mass $\mu=0.62$ and the lower limit on the lobe temperature of $1.6\times10^8$\,K \citep{WYK13}, we obtain $c_{\rm s}\ga1.9\times10^8$\,cm\,s$^{-1}$ ($\ga0.006c$). Using the lobe temperature of $2\times10^{12}$\,K, derived by \cite{WYK13} based on entrainment modelling and pressure constraints for the lobes\footnote{We have confirmed by considering the thermal bremsstrahlung emissivity of the giant lobes in the model of Wykes et al. that the {\it INTEGRAL} upper limits on photon counts from the giant lobes presented by \cite{BEC11} are at least two orders of magnitude above the predictions of the model for any lobe temperature.}, we can set an upper limit on the sound speed using the relativistic expression $c_{\rm s}=c/\sqrt{3}$ which yields $c_{\rm s}\la1.7\times10^{10}$\,cm\,s$^{-1}$ ($\la0.577c$). This gives us, taking the projected width of the {\it vertex} filament of $6.4$\,kpc, a timescale for its expansion in the range $t_{\rm cs}\sim37$\,kyr -- $3.3$\,Myr. For the {\it vortex} filament projected width of 3.6\,kpc, we obtain a timescale for expansion in the range $t_{\rm cs}\sim21$\,kyr -- $1.9$\,Myr. The derived timescales are possible to reconcile with the notion that these filaments, if plasma-dominated and mildly overpressured, are driven directly by a pre-existing jet given the limit on the age of the current jet of $\sim2$\,Myr (see \citealp{WYK13} and references therein).

If the filaments were more strongly overpressured (for example, if they are magnetically dominated as discussed in Section\,\ref{sect:pressure}) then the expansion speeds would be faster than the external sound speed and the timescales estimated above would be correspondingly reduced. However, our overall conclusion is that it is marginally possible for weakly overpressured filaments to have been produced directly by the last vestiges of former jet activity in the giant lobes.

\subsubsection{Internal MHD turbulence} \label{sect:turbulence}

Simulations (e.g. \citealp{CLA93, LEE03, TRE04, SCH04, FAL10, JON11, TEN13, HAR13}) have shown that magnetic filaments develop naturally as a consequence of MHD turbulence. The {\it vertex}, {\it vortex} and also the fainter, ribbon-like filaments in the southern giant lobe of Centaurus\,A are reminiscent of the outcomes of such simulations, and we conjecture that their origin could be due to the turbulent motions in the lobe, presumably driven by current or recently terminated\footnote{For a giant lobe disconnected from the energy supply, the timescale for decay of turbulence, with the parameters derived by \cite{WYK13}, is of order 6\,Myr.} jet activity \citep{WYK13}. Turbulence folds and stretches the $B$-field, leading to its amplification (dynamo action) and a spatially intermittent $B$-field distribution. Filaments develop with lengths of order of the largest eddies and transverse dimensions possibly as small as the viscous dissipation scale. $B$-fields coming via the turbulent dynamo concentrate into such structures. From the Jones et al. MHD simulations, the filaments are likely representing flux features stretched around the larger eddies in the turbulence. Even if the global helicity vanishes, some local `helicity' might be present that could instigate filament twisting. Weak shocks with a Mach number $\mathcal{M}\sim2$ are evident in density images from two-dimensional HD simulations (e.g. \citealp{LEE03}), and \cite{EIL14} alluded to the existence of transonic flows, and hence weak shocks, augmenting locally the Alfv\'enic turbulence, in Centaurus\,A's giant lobes. 

Filaments persist about an eddy turnover time for the scale of the eddies that stretch them; filament longevity goes as $\lambda^{2/3}$, which builds upon the Kolmogorov kinetic energy scaling ($E(k)\propto k^{-5/3}$, where $k=2\pi/\lambda$ is the wavenumber): velocity $v_{\rm \lambda}$ associated with scale $\lambda$ compares to that on the driving scale (i.e. outer turbulence scale) $v_{\rm \lambda max}$ as
\begin{equation}
v_{\rm \lambda}\simeq v_{\rm \lambda max}\,(\lambda/\lambda_{\rm max})^{1/3}\,, \label{eq:longevity}
\end{equation}
with $\lambda_{\rm max}$ the eddy size on the driving scale. However, the Kolmogorov scaling is not exactly in play when targeting the driving scale as processes on that scale take longer than on the inertial range, and so the filament longevity invoking $\lambda_{\rm max}$ might be considered a lower limit. To establish the eddy turnover time on the driving scale we use the turbulent speed $v_{\rm t}$ (i.e. $v_{\rm \lambda max}$) from \cite{WYK13}, $v_{\rm t}\sim 1.9\times10^{9}$\,cm\,s$^{-1}$, and a driving scale of $60$\,kpc (i.e. about a mid-value from their range $30-100$\,kpc); this gives $\sim3.1$\,Myr. Taking the {\it vertex}' and {\it vortex}' LAS (respectively $34$ and $53$\,kpc, see Section\,\ref{sect:introduction} and Table\,\ref{tab:parameters}) as a scale below $\lambda_{\rm max}$ and employing Equation\,\ref{eq:longevity}, then the speed at these scales is $v_{\rm \lambda=34kpc}\sim1.6\times10^9$\,cm\,s$^{-1}$ and $v_{\rm \lambda=53kpc}\sim1.8\times10^9$\,cm\,s$^{-1}$ and consequently their longevity, or `turbulent age', is respectively $\sim2.1$ and $\sim2.8$\,Myr. 

We can test the turbulence-generated model by considering the power spectrum of structure in the lobe, since a given magnetic field power spectrum will give rise to a characteristic power spectrum of projected synchrotron emissivity (e.g. \citealp{EIL89}) if we assume that the variations in electron energy density are small. Taking the two-dimensional Fourier transform of the 1.4\,GHz synchrotron emission (as the 1.4\,GHz image is the most sensitive available to us) and averaging its amplitude in radial bins to find the power on each spatial scale in the manner described by \cite{HAR13} we find that the power spectrum on the scales of the lobe that can be reliably measured (i.e. smaller than the largest scale of the lobe itself and larger than the resolution) is flatter than expected from a Kolmogorov spectrum for the $B$-field but there is no particular {\it a priori} reason to expect such a spectrum in real MHD turbulence; on the other hand, despite the subtraction of point sources from the image (see Section\,\ref{sect:14processing}) there is still a large amount of spurious small-scale structure which will affect the measured power spectrum. Overall, we can say that the turbulence model passes this consistency check, in the sense that the structure in the lobes does appear to have a power spectrum consistent with a power law; in particular, there is no evidence that the {\it vertex} and {\it vortex} are formed in a different way from the other filaments in the lobes, as would be provided by, e.g., a deviation from a power law on scales comparable to those structures.

It is difficult to put (independent) quantitative limits on the fundamental parameters of fully developed turbulence, such as the turbulent speed $v_{\rm t}$ at the driving scale\footnote{Wykes et al.'s (2013) estimate of $v_{\rm t}$ is based on the condition that a turbulent dynamo can reach energy equipartition, $U_B\sim U_{\rm t}$.}, or the kinematic viscosity $\nu$ and the corresponding viscous dissipation scale $\lambda_{\rm \nu}$, in an environment such as Centaurus\,A's lobes. The rarefaction and high temperature of the lobe plasma (in the model described by \citealp{WYK13}) imply that binary (Coulomb) collisions between ions (mainly protons) and/or electrons are too infrequent to be physically relevant. This makes the conventional collisional estimate for the kinematic viscosity of the form $\nu\simeq v_{\rm th,i}^2\,\tau_{\rm scat,i}/3\simeq v_{\rm th,i}^2/(3\,\nu_{\rm scat,i})$, with $v_{\rm th,i}=(3kT_{\rm i}/m_{\rm i})^{1/2}$ the ion thermal velocity, $\tau_{\rm scat,i}$ the ion-ion Coulomb collision time and $\nu_{\rm scat,i}$ the collisional frequency\footnote{The ion-ion collisional frequency is (e.g. \citealp{BRA65}) $\nu_{\rm scat,i}=4\pi^{1/2}Z^4n_{\rm i}e^4{\rm ln}\Lambda/(3m_{\rm i}^{1/2}k^{3/2}T_{\rm i}^{3/2})$, where ${\rm ln}\Lambda$ is the Coulomb logarithm in which $\Lambda=4\pi n_{\rm i} \lambda_{\rm D}^3/3$, and $\lambda_{\rm D}=(kT/4\pi n_{\rm i} e^2)^{1/2}$ is the Debye length.}, equally inappropriate. To illustrate: the collisional mean free path for proton-proton collisions in a Maxwellian plasma of temperature $T$ is of order $\lambda_{\rm mfp,i}\simeq v_{\rm th,i}\,\tau_{\rm scat,i}$. With a lower limit on the lobe temperature of $1.6\times10^8$\,K and a thermal proton number density $n_{\rm p, th}\sim 5.4\times10^{-9}$\,cm$^{-3}$ \citep{WYK13}, and thus $\tau_{\rm scat,i}\sim1.4\times10^{20}$\,s, $\lambda_{\rm mfp,i}\ga9$\,Gpc, i.e. larger than the source size. 
 
However, collective processes, in particular gyro-resonant interactions between protons and MHD waves or low-level MHD turbulence, likely produce an effective mean free path much smaller than the above estimate (see also, e.g., \citealp{SCH06, LAZ06, SAN14}). In the simplest model, with a low level of isotropic MHD turbulence with magnetic amplitude $\delta B$, pitch angle scattering by MHD waves (e.g. \citealp{WEN74}) restricts the mean free path along a large-scale magnetic field $B_{\rm 0}$ to $\lambda_{\rm mfp,\parallel}\simeq r_{\rm g,i}/(\delta B/B_{\rm 0})^2$ with $r_{\rm g,i}=v_{\rm th,i}/\Omega_{\rm i}$ the gyroradius of thermal ions and $\Omega_{\rm i}=ZeB_{\rm 0}/(m_{\rm i}\,c)$ the ion gyrofrequency.

There is some evidence that the power spectrum of MHD turbulence retains the Kolmogorov slope (e.g. \citealp{GOL95, GOLS95, CHO03, KIM05, KOW07, GAS13, CHO14}), with a flattening towards the driving scale. However, the `knee' (i.e. the transition from the flat-spectrum range to the Kolmogorov slope) is not well constrained. Observations provide a turbulence level $\delta B/B_{\rm 0}$ on the largest expanses, for which one generally adopts the fiducial value $\delta B/B_{\rm 0}\sim1$ (e.g. \citealp{SUL09}). A crude estimate of turbulence levels on small scales can be obtained by scaling down from the driving scale to a dimension that we determine invoking the relations $\delta B/B_0\propto\lambda\simeq(\lambda/\lambda_{\rm max})^{1/3}$ and $\lambda_{\rm mfp,\parallel}\simeq r_{\rm g,i}/(\delta B/B_0)^2$, and assuming $B_0$ similar on both scales. The mean free path derived in this way gives an upper limit on the true value since the magnetic field power spectrum may flatten towards the largest scales. Working from the driving scale of $\lambda_{\rm max}=60$\,kpc as above we arrive at $\lambda_{\rm mfp,\parallel}\la r_{\rm g,i}^{3/5}\,\lambda_{\rm max}^{2/5}\la3.4\times10^{15}$\,cm ($\la1\times10^{-3}$\,pc); such a proton effective mean free path is fairly below the size of the resolved fine structure in Centaurus\,A's giant lobes. At this scale, the derived turbulence level is $\delta B/B_0\la2.6\times10^{-3}$. This leads to an effective viscosity where large-scale motions (on scales significantly larger than $\lambda_{\rm mfp,\parallel}$) with velocity $v$ are dissipated with a dissipation rate per unit volume $\dot{\varepsilon}$. If scattering is weak ($\nu_{\rm scat,i}\ll\Omega_{\rm i}$), the dissipation is mostly due to compression and motion along the magnetic field, taken to be the $z$-direction:
\begin{equation}
\dot{\varepsilon}=n_{\rm i}m_{\rm i}\nu_{\rm eff}\,\Big(\frac{\partial v_{\rm z}}{\partial z} - \frac{1}{3} \nabla \cdot {\emph{\bf v}} \Big)^2\,,
\label{eq:diss}
\end{equation}
see, e.g., \cite{KAU60} and \cite{BRA65}.
The above assumes the effective collision time $\tau_{\rm scat,i}\simeq \lambda_{\rm mfp,\parallel}/v_{\rm th,i}\simeq 1/(\Omega_{\rm i}\,(\delta B/B_{\rm 0})^2)$. Note that the dissipation rate scales as $\dot{\varepsilon}\propto n_{\rm i}m_{\rm i}\nu_{\rm eff}\,v_{\rm t}^2(\lambda)/\lambda^2$
with $\lambda$ the turbulent scale and $v_{\rm t}(\lambda)$ the turbulent speed at that scale;
this scaling with $v_{\rm t}$ and $\lambda$ is the same as for ordinary (collisional) viscosity. 

The purpose of the next three steps is to compute the scales at which the turbulent cascade dissipates: the viscous cutoff that marks the end of the HD structure evolution, and the resistive cutoff that marks the end of the MHD architecture, to finally obtain a crude estimate of the current sheet dimension likely associated with the resistive scale. This allows us to constrain the growth time of the tearing mode.
  
In a simple Krook collision model \citep{BHA54} with parallel scattering mean free path $\lambda_{\rm mfp,\parallel}$ and thermal velocity $v_{\rm th,i}$, the effective kinematic viscosity entering Equation\,\ref{eq:diss} is
\begin{equation}
\nu_{\rm eff}\simeq v_{\rm th,i}\,\lambda_{\rm mfp,\parallel}\,/\,15\,. 
\label{eq:nueff}
\end{equation}
Adopting $T=1.6\times10^8$\,K, $B=0.9$\,$\mu$G and $\delta B/B_{\rm 0}=2.6\times10^{-3}$, we get $\nu_{\rm eff}\sim4.5\times10^{22}$\,cm$^2$\,s$^{-1}$; this is considerably smaller than the kinematic viscosity based on Coulomb collisions. Using such effective viscosity, the Reynolds number is, adopting the driving scale and turbulent speed figures as above, Re$\,=\lambda_{\rm max}\,v_{\rm t}/\nu_{\rm eff}\sim7.7\times10^9$, which implies that a Kolmogorov cascade breaks off at $\lambda_{\rm \nu}\simeq \lambda_{\rm max}/{\rm Re}^{3/4}\sim7.1\times10^{15}$\,cm ($\sim2\times10^{-3}$\,pc). Thus, in principle, a turbulent cascade could be set up from kpc to sub-pc scales comparable to $\lambda_{\rm mfp,\parallel}$, assuming $T$, $B$ and $\delta B/B_{\rm 0}$ levels as above. 

The magnetic diffusivity due to electron scattering with scattering time $\tau_{\rm scat,e}$, $\eta_{\rm m}=c^2/4\pi\sigma$, with $\sigma=n_{\rm e}\,e^2 \tau_{\rm scat,e}/m_{\rm e}$ the electrical conductivity, is expected to be $\ll\nu_{\rm eff}$ for typical parameters. In this situation magnetic dynamo action is allowed at scales $\ll$ the dissipation scale of the large-scale kinetic turbulence (e.g. \citealp{TOB11}). An actual computation of $\eta_{\rm m}$ is hampered by the same problem as calculating the effective viscosity: the theory of Coulomb collisions, in this case for the current-carrying electrons, does not apply as the electron mean free path is comparable to the source size. We parametrize this uncertainty by expressing all parameters in terms of the (unknown) electron scattering mean free path $\lambda_{\rm mfp,e}$ in the turbulent medium. For thermal electrons, the scattering time is $\tau_{\rm scat,e}\simeq\lambda_{\rm mfp,e}/(kT/m_{\rm e})^{1/2}$ and the magnetic diffusivity becomes
\begin{equation}
\eta_{\rm m} = \frac{(c/\omega_{\rm pe})^2}{\tau_{\rm scat,e}}\,,
\end{equation}
where $c/\omega_{\rm pe}$ is the electron skin depth and $\omega_{\rm pe}= (4\pi n_{\rm e}e^2/m_{\rm e})^{1/2}$ the electron plasma frequency.

The magnetic Reynolds number Re$_{\rm m}$ for large-scale fields is based on the magnetic diffusivity (as above) $\eta_{\rm m}=c^2/4\pi\sigma$, with here $\sigma=\omega_{\rm pe}^2 \tau_{\rm scat,e}/4\pi$ the conductivity of the plasma. Using for the scattering time the anomalous (i.e. turbulence-induced) value $\tau_{\rm scat,e}\simeq\lambda_{\rm mfp,\parallel}/v_{\rm th,e}$ with the bulk of the electrons (since they carry the current), we have $r_{\rm g,e}=r_{\rm g,i}(m_{\rm e}/m_{\rm i})^{1/2}\sim5.4\times10^8$\,cm ($\sim2\times10^{-10}$\,pc) and $\tau_{\rm scat,e}=\tau_{\rm scat,i}(m_{\rm e}/m_{\rm i})^{1/2}\sim9.4\times10^3$\,s. For $B=0.9$\,$\mu$G and $\delta B/B_0=2.6\times10^{-3}$, this yields an anomalous conductivity $\sigma\sim1.3\times10^4$\,s$^{-1}$. The specific magnetic diffusivity is then $\eta_{\rm m}\sim5.6\times10^{15}$\,cm$^2$\,s$^{-1}$ which is smaller than the effective kinematic (ion) viscosity. This means that the magnetic Prandtl number Pr$_{\rm m}$, defined as Pr$_{\rm m}\equiv\nu_{\rm eff}/\eta_{\rm m}\sim8.1\times10^6$, is large; therefore magnetic turbulence can be maintained well below the viscous dissipation scale $\lambda_{\nu}$. However, the character of the turbulence is different on those scales, since (both fast and slow) magnetosonic fluctuations are damped, velocity fluctuations are supposedly far sub-Alfv\'enic and the $B$-field lines are mostly `shuffled'. Relating $\lambda_{\nu}$ to the magnetic Prandtl number, one finds the resistive dissipation scale, $\lambda_{\eta}=\,$Pr$_{\rm m}^{-1/2}\lambda_{\nu}\sim2.5\times10^{12}$\,cm ($\sim8\times10^{-7}$\,pc). This is well below any observable scale, but we will re-appeal to the resistive cutoff in Section\,\ref{sect:instabilities} in the context of the tearing mode.

\subsubsection{(Surface) instabilities} \label{sect:instabilities}

We do not as yet have a tight observational constraint on whether the {\it vertex} and/or {\it vortex} filaments are surface features or are fully embedded in the lobe. An origin as magnetic flux ropes created by surface instabilities, such as KH or RT, is therefore somewhat appealing. \cite{FEA11} have suggested the KH instability to be at the origin of some of the filamentary features in the southern lobe, albeit without detailed arguments or support from modelling. \cite{WYK13} and \cite{EIL14} have argued that the giant lobes are in an approximate pressure balance with their surroundings (a prerequisite for the KH instability to operate) and we further verify whether the KH instability could develop by considering its growth time (e.g. \citealp{CHA61}): 
\begin{equation}
t_{\rm KH} = \Big[\,\frac{(\rho_{\rm l} + \rho_{\rm g})^2}{k^2\rho_{\rm l}\,\rho_{\rm g}\,v_{\rm l-g}^2}\,\Big]^{1/2}\,, \label{eq:KH}
\end{equation}
where $k$ is the wavenumber, $\rho_{\rm l}$ and $\rho_{\rm g}$ the giant lobe and intragroup plasma mass densities and $v_{\rm l-g}$ the velocity of the lobe flow relative to the intragroup flow. Taking 5\,kpc for the scale (that is, the mean filament thickness, see Table\,\ref{tab:parameters}) and thus $\sim4.1\times10^{-22}$\,cm$^{-1}$ for the wavenumber, $1.1\times10^{-8}$\,cm$^{-3}$ \citep{WYK13} and $1\times10^{-4}$\,cm$^{-3}$ \citep{SUL13, EIL14} for respectively the total lobe and intragroup densities, and the buoyancy velocity of $4.9\times10^7$\,cm\,s$^{-1}$ ($\sim0.002c$) \citep{WYK13} for the lobe flow speed (while setting the intragroup speed to zero), we obtain a growth rate of $\sim4.1\times10^{-9}$\,yr$^{-1}$. From this it follows that the growth time is of order $t_{\rm KH}\sim240$\,Myr, which makes the origin of the {\it vertex}/{\it vortex} as a KH instability improbable. If, however, the interior plasma is moving faster than the buoyancy speed at which we propose the giant lobes are rising and/or the thermal particle content of the lobe is higher than assumed here, the growth time will be smaller and an explanation in terms of KH instability less restrictive. In the limit that $n_{\rm th}\sim1\times10^{-4}$\,cm$^{-3}$ \citep{STA13, SUL13}, the KH growth time is $\sim3.4$\,Myr. In any case, the origin as a surface KH instability may be ruled out observationally given the fact that we do not see perturbations at the (projected) edges of the lobes with sizes comparable to the {\it vertex} or {\it vortex}.

The RT instability is driven by buoyancy in a stratified plasma and requires heavy material on top of lighter matter. It is not clear that the effect of RT (`fingers' of external medium intruding into the lobes) could give rise to structures resembling the {\it vertex} or {\it vortex}.

The RM instability is analogous to the RT mode. It requires low-density material intruding impulsively into a higher-density matter, or vice versa, producing voids in the former case and spike-like features in the latter. The RM instability is probably more relevant for very young lobes; we have no clear indication at such RM-like features in the available radio images of the filaments, and we do not expect strong shocks at the distance from the core of the {\it vertex}/{\it vortex} filaments in the giant lobes (see, e.g., Sections\,\ref{sect:pressure} and \ref{sect:jet}). 

The tearing instability develops due to small non-zero resistivity in high-current regions with $B$-field reversals. The primary requirement for this instability to be triggered is a thin field-reversal layer (the geometry as in magnetic reconnection), which could occur at the lobe/intragroup medium interface and probably also inside the lobes. We can legitimately consider anomalous resistivity, i.e. resistivity arising from particle-wave interactions (see, e.g., \citealp{LAV66}; \citealp{MEL94} and references therein; Section\,\ref{sect:turbulence}). We can adopt an extreme upper limit on the anomalous collision rate (following \citealp{HIN89} in their treatment of the tearing mode possibly associated with lobe filaments in M\,87) which is the electron plasma frequency $\omega_{\rm pe}=(4\pi\,n_{\rm e}\,e^2/m_{\rm e})^{1/2}$, and so the collision rate is $\nu_{\rm scat,e}\la\omega_{\rm pe}/2\pi$. This gives us an upper limit on the resistivity: $\eta\la2c^2/\omega_{\rm pe}$. A lower bound on the growth time for the tearing mode then reads \citep{HIN89}
\begin{equation}
t_{\rm TM} \ga \frac{(2\pi\,\omega_{\rm pe})^{3/5}\,a^{8/5}}{(c^3\,v_{\rm A})^{2/5}}\,, \label{eq:TM}
\end{equation}
in which $a$ denotes the width of the current sheet and $v_{\rm A}$ the Alfv\'en speed. We use $v_{\rm A}=2.4\times10^9$\,cm\,s$^{-1}$ ($\sim0.081c$) and the thermal electron content $n_{\rm e,th}=5.4\times10^{-9}$\,cm$^{-3}$ \citep{WYK13} as an approximation to the total electron number density (it is the full plasma which carries the waves). The width of the current sheet is not well known, yet an obvious hard lower limit is the electron gyroradius which is for our adopted conditions in the giant lobes $r_{\rm g,e}\sim5.4\times10^8$\,cm ($\sim2\times10^{-10}$\,pc). Probably the most realistic estimate of the current sheet width\footnote{In the presence of MHD turbulence, a single stochastic magnetic reconnection region is associated with a multitude of such current sheets (e.g. \citealp{LAZ99}). Magnetic reconnection and weak shocks associated with the filaments could locally boost the Alfv\'enic particle acceleration likely operating throughout the lobes \citep{EIL14}.} is the resistive dissipation scale of the turbulence that we have estimated in Section\,\ref{sect:turbulence} to be $\lambda_{\eta}\sim2.5\times10^{12}$\,cm ($\sim8\times10^{-7}$\,pc). This gives $t_{\rm TM}\ga6.3$\,h, which obviously falls well within both the dynamical and spectral ages of the lobes, and also within the turbulent ages of the individual {\it vertex} and {\it vortex} filaments (derived in Section\,\ref{sect:turbulence}). The resistive tearing mode is generally suppressed in high Re number media; however, the eddy turnover time at the tearing scale ($\sim0.2$\,yr) is longer than the lower limit on the tearing instability growth time, hence the tearing is presumably not suppressed by this route. The topological consequences of the tearing instability are tiny closed $B$-field loops in 2D simulations and long magnetised filaments in 3D. Thus, while the tearing instability may be associated with the {\it vertex} and {\it vortex} filaments, it would be expected to show a non-{\it vertex}/{\it vortex} morphology and to appear on unobservable scales within them.

Various authors considered radiative instabilities, which arise as a result of runaway cooling, in radio galaxies' lobes: synchrotron cooling, if relativistic electrons dominate the pressure \citep{HIN89, DEG89, BOD90, ROS93}, or thermal bremsstrahlung cooling, if thermal plasma dominates \citep{HIN89, BOD90}. To deal with the synchrotron instability, we consider relativistic electrons with a distribution of energies $n_{\rm e,rel}(E_{\rm e})=N_0\,E_{\rm e}^{-p}$ between $E_{\rm e,min}$ and $E_{\rm e,max}$, where $N_0$ is the normalisation of the electron energy spectrum and $p$ is the electron spectral index. The growth time of the synchrotron instability, with also the inverse-Compton component included, can be written as the total electron energy divided by the total electron energy loss rate:
\begin{equation}
t_{\rm sync}=\frac{\int_{E_{\rm e,min}}^{E_{\rm e,max}}n_{\rm e,rel}(E_{\rm e})\,E_{\rm e}\, {\rm d}E_{\rm e}}{C \int_{E_{\rm e,min}}^{E_{\rm e,max}}n_{\rm e,rel}(E_{\rm e})\,E_{\rm e}^2\, {\rm d}t}\,,
\end{equation}
where $C$ is a constant such that for an electron $ $d$E_{\rm e}/$d$t=-CE_{\rm e}^2$, and it represents both synchrotron and inverse-Compton losses: $C=4\sigma_{\rm T}\,(U_{B{\rm age}}+U_{\rm CMB})/(3m_{\rm e}^2\,c^3)$, where $\sigma_{\rm T}$ denotes the Thomson cross section, $U_{B{\rm age}}$ the energy density in aged $B$-field (as in Section\,\ref{sect:ages}) and $U_{\rm CMB}$ the energy density in CMB photons (see also \citealp{HAR13}). With $B=0.9$\,$\mu$G, $p=2$, and $E_{\rm e,min}=5$\,MeV and $E_{\rm e,max}=100$\,GeV (which we have invoked in calculations of the filaments and the giant lobes) as energy cutoffs, the instability growth time becomes $\sim110$\,Myr. This exceeds the turbulent ages of the individual filaments as well the spectral ages of the lobes. The synchrotron instability may be also excluded on grounds of the likely dominance of non-radiating particle pressure in the lobes (\citealp{WYK13} for Centaurus\,A; e.g. \citealp{CRO14} for other FR\,I sources). The growth time of the thermal bremsstrahlung instability goes as (e.g. \citealp{KAR61})
\begin{equation}
t_{\rm th-br}= 1.8\times10^{11}\,\frac{T^{1/2}}{n_{\rm e,th}\,\bar{g}_{\rm ff}}\,, 
\end{equation}
where $T$ is the lobe temperature and $\bar{g}_{\rm ff}$ the temperature-averaged Gaunt factor. Other thermal energy and emissivity parameters are absorbed into the numerical prefactor. Using the range of temperatures $1.6\times10^8$ to $2.0\times10^{12}$\,K and $n_{\rm e,th}=5.4\times10^{-9}$\,cm$^{-3}$ as above, and setting $\bar{g}_{\rm ff}=10$ \citep{KAR61}, we obtain a growth time range $\sim 1.3-99.6$\,Pyr which is much larger than the Hubble time. Even if the thermal density were as high as $1\times10^{-4}$\,cm$^{-3}$ and temperature as low as $5.8\times10^6$\,K (\citealp{STA13}), the thermal bremsstrahlung instability would still exceed the Hubble time, as expected. This makes a development of radiative instabilities in the giant lobes unrealistic.

The current-driven instabilities (CDI) do not as much address the question of the origin of the filaments but rather the question what happens to them once formed; we will treat two such instabilities below.

The sausage instability, formally an axisymmetric $m=0$ mode in a cylindrical magnetised
plasma column (pinch), occurs when the azimuthal field $B_{\rm \phi}$ becomes too strong. In that case a sausage-like series of bulges and compressions in the tube radius grow; the exact growth rate of this mode depends on the field geometry both inside the pinch and in the surrounding medium. Filaments in other radio lobes, where measured, usually show a relatively high degree of linear polarization, implying the presence of significant axial field $B_{\rm z}$ (e.g. Pictor\,A, \citealp{PER97}; 3C\,310, \citealp{BRE84}; MG\,0248+0641, \citealp{CON98}; Hercules\,A, \citealp{DRE84, SAX02, GIZ03}). Also, the \cite{JUN93} polarization images of Centaurus\,A seem to show $E$-field vectors perpendicular to the $z$-axis of the filaments (albeit the resolution of these maps is not very high). This makes the sausage instability as a means of affecting the {\it vertex}/{\it vortex} shape rather unlikely; moreover, inspecting in more detail the radio images, the {\it vertex} and {\it vortex} morphology does not well match the morphology as expected for the sausage mode.

The kink instability ($m=1$ mode) eventuates when a plasma column moves more-or-less as a whole, and its axis is deformed into a sinusoidal or helical shape. Analogously to the sausage mode, an exact determination of growth rates and the stability criterion depends on the detailed properties of the magnetic field inside and outside the pinch; the presence of an azimuthal field is destabilising. The instability develops when the magnetic twist $\Delta \phi$ exceeds a critical value\footnote{\cite{HOO81} derived a critical value of 2.49$\pi$. However, this outcome is model-dependent and cannot be considered universally.} (\citealp{LUN51, HOO79, HOO81, KLI04}). The evolution of the kink instability has been investigated by, e.g., \cite{BEG98} and \cite{HAY08}, and \cite{SRI10} have provided observational evidence for the $m=1$ mode in the solar corona. The growth time of the kink instability can be approximated as (e.g. \citealp{BAT78})
\begin{equation}
t_{\rm m=1}=\frac{r}{v_{\rm A}}\,\Big[\,\frac{1}{k^2\,r^2 - (B_{\rm \phi}(r)/B_{\rm z})^2}\,\Big]^{1/2}\,, \label{eq:m=1}
\end{equation} 
where $k$ is again the wavenumber, $r$ is the flux tube radius and $B_{\rm \phi}(r)$ the azimuthal field at the pinch surface (i.e. the edge of the filament). In the available radio images, the kink mode is mostly matched by the {\it vertex} morphology, hence we will calculate the growth time for this particular filament. We consider the northern part of the {\it vertex} which seems to show a turn over >1 wavelength (see Figs.\,\ref{fig:fig2} and \ref{fig:fig3}). The {\it vertex} radius is (at the widest point) 3.6\,kpc; as before, we adopt $k=4.1\times10^{-22}$\,cm$^{-1}$ and $v_{\rm A}=2.4\times10^9$\,cm\,s$^{-1}$, and $B_{\rm \phi}/B_{\rm z}\sim0.5$ (from Section\,\ref{sect:twist}). This results in a growth time of $\sim32$\,kyr. This is, as for the tearing mode, within both the dynamical and spectral ages of the lobes and also within the turbulent ages of the {\it vertex} and {\it vortex} filaments. A co-temporary development of the kink and the tearing mode, on disparate scales, is possible. Thus, the kink instability is likely to be operating: not destroying the {\it vertex} filament, just making it appear kinked.

\subsubsection{Dwarf galaxy KK\,196 (AM\,1318--444)} \label{sect:galaxy}

We consider explanation (iv) in Section\,\ref{sect:origin} the least likely given the low mass of the dwarf irregular galaxy KK\,196, its low relative velocity to NGC\,5128, a combination of high temperature and low particle density of the giant lobe plasma, and the {\it vertex} and {\it vortex} morphology; moreover, we would expect a single wake. We first test this assertion by performing a simple check for a galaxy to have a {\it Bondi-Hoyle wake}: the accretion radius $R_{\rm acc}$ should be larger than its size. The accretion radius is given by $R_{\rm acc} = 2 G M_{\rm gal}/(v_{\rm gal}^2 + c_{\rm s}^2)$, where $M_{\rm gal}$ and $v_{\rm gal}$ are the mass and the velocity of the traversing galaxy, and $c_{\rm s}$ the local speed of sound (\citealp{BON52, SAK00}). Adopting $15$\,arcsec as KK\,196's angular radius \citep{JER00} we infer $\sim1.8\times10^{21}$\,cm for its physical size. 

The total baryonic mass of KK\,196 is $\le5.9\times10^7$\,M$_{\odot}$ (based on the results from \citealp{JER00} and \citealp{WAR07}), and its relative radial velocity to NGC\,5128 is approximately 189\,km\,s$^{-1}$, so we estimate a relative 3D velocity of $v_{\rm gal}\sim189\times\sqrt{3}\sim327$\,km\,s$^{-1}$ ($\sim0.001c$). Note that \cite{CRN12} derive KK\,196's total baryonic mass as $5.1-7.2\times10^7$\,M$_{\odot}$ based on its $B$-band luminosity, a stellar mass-to-light ratio $1<M/L<2$ and the measured H\,I mass, and as $6.9-8.2\times10^7$\,M$_{\odot}$ based on KK\,196's star formation history and the measured H\,I mass. The baryonic mass is thus well constrained and within $5-8\times10^7$\,M$_{\odot}$.

Assessing the non-baryonic dark matter (DM) mass fraction in dwarf irregular galaxies is difficult because the rotation curves are generally hard to measure, the H\,I rotation velocities increase only very slowly as a function of galactocentric distance and peak typically at an amplitude not much larger than the turbulent motion of the gas. From a few known cases (see \citealp{COT00}) follows that the DM mass is of the same order as the stellar mass. For KK\,196 this translates into a total mass $M_{\rm tot}$ of $4.5\times10^7<M_{\rm stellar}+M_{\rm gas}+M_{\rm DM} <7.3\times10^7$\,M$_{\odot}$. Note that the dwarf irregular ESO\,444-G084 in the C\^ot\'e et al. sample has a luminosity comparable to KK\,196 and its $M_{\rm DM}/M_{\rm lum}$ value at $R_{25}$ (radius at the 25th mag/arcsec$^2$ surface brightness) is $0.3-0.5$. Thus, with the estimate given above we are likely to slightly overestimate KK\,196's total mass. However, they also quote $M_{\rm DM}/M_{\rm lum}=10-12.9$ for ESO\,444-G084 at $R_{\rm max}$. If the comparison between KK\,196 and ESO\,444-G084 holds out to that larger radius then the total mass of KK\,196 is of the order of $M_{\rm tot}\sim5\times10^8$\,M$_{\odot}$\footnote{For dwarf galaxies, a significant increase in non-baryonic DM is common. The $R_{25}$ radius is roughly the size of the stellar component of the galaxy. However, the H\,I-based rotation curves tend to flatten out onliest further out (e.g. \citealp{WAR06}), hence there might be a vast amount of non-baryonic DM in these systems beyond their optical radii (see also, e.g., \citealp{SIM07}).}. This value is also consistent with $M_{\rm tot}=M_{\rm dyn}=1-4\times10^8$\,M$_{\odot}$, where $M_{\rm dyn}$ is the dynamical mass, inferred from \cite{EDE00} adopting $M_B=-12.16$ and a colour $B-I=1.5$ (i.e. total $I$-band luminosity of $-13.66$) for KK\,196. 

In the non-relativistic limit and assuming $T\ge1.6\times10^8$\,K \citep{WYK13}, the local sound speed in the giant lobes reaches $\ga1.9\times10^8$\,cm\,s$^{-1}$ ($\ga0.006c$; see also Section\,\ref{sect:jet}) and consequently $R_{\rm acc}\la3.6\times10^{18}$\,cm. Hence, the accretion radius is at least $3$ orders of magnitude smaller than KK196's size and no Bondi-Hoyle wake is expected. If the internal sound speed in the lobes is relativistic (as is required in any scenario where there is stochastic acceleration of ultra-high energy cosmic rays) then the dwarf galaxy has even less effect.  

Another possible manifestation of the galaxy-lobe interaction is a bow shock in front of the traversing galaxy and associated stripping of the galaxy ISM: then the filaments may be associated with a ram pressure-induced wake. This however requires KK\,196's velocity to be in excess of the intralobe sound speed which is not the case (see above). Still a moderate gas stripping could occur from KK\,196 in the absence of a bow shock: the non-detection of H\,I \citep{BAN99} makes gas removal from KK\,196 at some point in the past plausible. However, this stripped-off H\,I gas is not likely to organize itself into filaments of the morphology of the {\it vertex}/{\it vortex} inside the giant lobes, and the H\,I nature is ruled out by the ATCA and our GMRT continuum observations.

As it has propagated, the current or former jet could have impacted on the dwarf galaxy KK\,196 with the {\it vertex}/{\it vortex} filaments as a result, but in objects where we actually see a jet interacting with a galaxy \citep{EVA08}, the results seem rather different, therefore we do not consider this scenario realistic. Moreover, this stripped-off H\,I-enriched gas would likewise not appear in the radio continuum images of the lobes.

\section{Summary and Conclusions} \label{sect:summary}

We have presented new, high-dynamic range GMRT observations at 325 and 235\,MHz of parts of the southern giant lobe of Centaurus\,A, and have modelled the origin of the filamentary structure associated with the lobe. The key results of this paper are as follows:

1. The detection at 325 and 235\,MHz, with comparable resolution to the 1.4\,GHz ATCA images, confirms the reality of the {\it vertex} and {\it vortex} filaments associated with the southern giant lobe. The spatial extent of the {\it vertex} and {\it vortex} nearly coincides with their morphology at 1.4\,GHz, reinforcing their synchrotron origin. The {\it vertex} shows fine substructure and appears twisted. The {\it vertex} and {\it vortex} fields demonstrate surplus filamentary features, however, we find no clear connection at 325\,--235\,MHz of the filaments to the extant or extinct jet. 

2. Combining the ATCA and GMRT data, and restricting the range of baselines to $0.15-2.5$\,k$\lambda$, we have inferred a spectral index $\alpha=0.81\pm0.10$ for the {\it vertex} filament and $\alpha=0.83\pm0.16$ for the {\it vortex}. This is marginally steeper than the spectral index of the general Centaurus\,A lobe plasma, and we discuss this in terms of an excess-loss scenario with particle confinement in the filaments. Our spectral fitting implies spectral ages that are difficult to reconcile with the `turbulent ages' (i.e. the longevity of the filaments stretched by the turbulent eddies) of $\sim2-3$\,Myr without invoking very high magnetic field strengths.

3. Our minimum pressure analysis for the filaments reveals $1.5\times10^{-13}$\,dyn\,cm$^{-2}$ ({\it vertex}) and $1.3\times10^{-13}$\,dyn\,cm$^{-2}$ ({\it vortex}) which is a factor $\sim3$ higher than the minimum global pressure of the lobes. Such an overpressure could be identified with weak shocks (Mach number $\mathcal{M}\sim1.7$). Synchrotron emissivity ratios result in a pressure jump of about $1.3$, compatible with expansion at the sound speed. No efficient Fermi\,I-type particle acceleration is expected for such slow expansions, and it also makes a spherical cocoon collapsing scenario with $\mathcal{M}\gg1.7$ in the giant lobes untenable.

4. Scaling down from the driving scale in the giant lobes with a presumable turbulence level of $\delta B/B_0\sim1$, we derive a turbulence level $\delta B/B_0\ga3\times10^{-3}$ at $\sim10^{-3}$\,pc. The viscous scale in the giant lobes is clearly considerably smaller than that derived from Coulomb scattering. From our calculated low value of the specific magnetic diffusivity, and therefore a large magnetic Prandtl number, we deduce that the MHD turbulence extends well below the viscous dissipation scale, with the resistive dissipation scale of order $10^{-6}$\,pc. This sets the width of the current sheet of the tearing mode.

5. We have considered several mechanisms that may generate the filaments in the southern giant lobe. We have shown that the dwarf irregular galaxy KK\,196 (AM\,1318--444) can be excluded as the origin of the {\it vertex}/{\it vortex} filaments, principally on grounds of its low relative velocity and mass, and on the morphology of the {\it vertex}/{\it vortex}. Our turbulence power spectrum modelling does not support different origin of the {\it vertex} and/or {\it vortex} from the other filaments in the lobe. The Kelvin-Helmholtz instability provoked at the lobe-intragroup interface is ruled out based on its growth time and observational constraints. The Rayleigh-Taylor and Richtmyer-Meshkov instabilities are highly unlikely to account for the {\it vertex}/{\it vortex} within the available observational constraints. Of the MHD instabilities, the tearing mode is supported by its growth time which lies conveniently within both the dynamical and spectral ages of the lobes and also within our derived turbulent ages of the {\it vertex}/{\it vortex}. The filaments are inconceivably affected by the sausage instability given the presumable presence of an axial $B$-field in the {\it vertex}/{\it vortex} which suppresses its growth. To the contrary, the kink instability {\it is} supported by a presence of such axial field, moreover, the existence of the kink instability is anticipated based on the filament radio morphology and on the growth time which is again sufficiently within the dynamical and spectral ages of the lobes and within the turbulent ages of the {\it vertex}/{\it vortex}. The radiative instabilities are ruled out based on their growth times. We lean towards the {\it vertex} and {\it vortex} filaments originating from intralobe MHD turbulence or from last stages of the activity of the pre-existing jet, or an interplay of both. 

There are several aspects that remain to be explored. The magnetic field direction and a possible variability of the spectral index along the filaments may be revealed by observations with the ATCA-CABB at GHz frequencies. With an instantaneous field of view of $30$\,deg$^2$ between 700 and 1800\,MHz and $20$ times higher spatial resolution in comparison to the current ATCA and GMRT images, the Australian SKA Pathfinder (ASKAP) will be the ideal facility for further investigations of the origin of the filaments, including the faint edge-like features, the {\it wisps}. In a future paper we will report on {\it XMM-Newton} observations designed to constrain the magnetic field strength of the individual filaments. Reproducing the complex morphology of the {\it vertex} will be a challenge for future MHD simulations.

\section*{Acknowledgments}
We thank the staff of the GMRT who have made these observations possible. GMRT is run by the National Centre for Radio Astrophysics of the Tata Institute of Fundamental Research. This research has made use of the NASA/IPAC Extragalactic Database (NED) which is operated by the Jet Propulsion Laboratory, California Institute of Technology, under contract with the National Aeronautics and Space Administration. We are indebted to S. Sirothia for providing the GMRT flux correction factors. SW thanks J.\,Eilek, P.\,Biermann, C.\,Elenbaas, P.\,Prasad, I.\,Feain, J.\,Bray, B.\,Marcote, J.\,Croston, L.\,Houben, F.\,Israel, F.\,Owen and J.\,P.\,Leahy for stimulating discussions, and an anonymous referee for a constructive report. TWJ gratefully acknowledges support from NSF grant AST1211595 and NASA grant NNX09AH78G.

\bsp

\label{lastpage}

\end{document}